\RequirePackage{amsmath}
\documentclass[12pt]{iopart}
\pdfoutput=1 
\usepackage{graphicx}
\usepackage{amsmath,amsfonts,amssymb}

\usepackage{xcolor}
\colorlet{mylinkcolor}{blue!66!black!80}
\usepackage[colorlinks=true,linkcolor=mylinkcolor,citecolor=mylinkcolor,filecolor=cyan,urlcolor=mylinkcolor,breaklinks=true]{hyperref}

\usepackage[utf8]{inputenc}
\usepackage{cite}

\newcommand{\f}[1]{\mathbf{#1}}
\newcommand{\om}{\omega}
\newcommand{\abs}[1]{\left\lvert #1 \right\rvert}
\newcommand{\defas}{\equiv}
\newcommand{\variance}{V}
\newcommand{\Q}{{\rm Q}}
\newcommand{\A}{{\rm A}}
\newcommand{\utof}{{\rm u \to\rm f}}
\newcommand{\utauf}{{{\rm u}\overset\tau\to\rm f}}

\begin{document}
\title{Scattering fingerprints of two-state dynamics}
\author{Cai Dieball$^1$, Diego Krapf$^{2}$, Matthias Weiss$^3$, Alja\v{z} Godec$^1$}
\address{$^1$ Mathematical bioPhysics Group, Max Planck Institute for
  Biophysical Chemistry, G\"ottingen, Germany}
\address{$^2$ Electrical and Computer Engineering and School of
  Biomedical Engineering, Colorado State University, Fort Collins,
  Colorado, USA}
\address{$^3$ Experimental Physics I, University of Bayreuth, Bayreuth, Germany}
\ead{agodec@mpibpc.mpg.de}

\begin{abstract}
Particle transport in complex environments such as the interior of living cells is often (transiently) non-Fickian or anomalous, that is, it deviates
from the laws of Brownian motion. Such anomalies may be the result
of small-scale spatio-temporal heterogeneities in, or viscoelastic
properties of, the medium, molecular crowding, etc. 
Often the observed dynamics displays
multi-state characteristics, i.e.\ distinct modes of transport dynamically
interconverting between each other in a stochastic
manner. Reliably distinguishing between single- and multi-state dynamics is
challenging and requires a combination of distinct approaches. To complement the
existing methods relying on the analysis of the particle's mean
squared displacement, position- or displacement-autocorrelation
function, and propagators, we here focus  on ``scattering
fingerprints'' of multi-state dynamics.  
We develop a theoretical framework for two-state
scattering signatures -- the intermediate scattering function and dynamic structure
factor -- and apply it to the analysis of simple model
systems as well as
particle-tracking experiments in living cells.
We consider inert tracer-particle motion
as well as systems with an internal structure and dynamics. 
Our results may generally be relevant for the
interpretation of state-of-the-art differential dynamic microscopy
experiments on complex particulate systems, as well as inelastic or
quasielastic neutron (incl.\ spin-echo) and X-ray scattering
scattering probing structural and dynamical properties of
macromolecules, when the underlying dynamics displays two-state transport.
\end{abstract}
\noindent{\it Keywords\/}: dynamic structure
factor, intermediate scattering function, diffusion, fractional Brownian motion, Gaussian
network model, segmentation

\maketitle

\section{Introduction}\label{sec:intro}
Transport in complex systems such as disordered media \cite{Bouchaud1990PR}, biological cells
\cite{Eli_PhysToday,Franosch2013}, and cell membranes \cite{Krapf_2015,Ilpo}, frequently
deviates from the laws of Brownian motion that govern the dynamics of
particles in simple media and at high dilution. The reason for the
deviation from the paradigmatic Brownian behavior may be sought in
spatial obstruction and/or macromolecular crowding
\cite{Eli_PhysToday,Franosch2013,Ilpo}, viscoelastic properties of the
medium \cite{Guigas_2008,McKinley_2009,Lene,Weitz,Godec_2014}, and spatial heterogeneities \cite{Slater,Alexei}, to name but a
few. Over the years a multitude of approaches have emerged to
characterize anomalous diffusion from different perspectives, and in
particular to distinguish between different modes of anomalous
diffusion (for excellent reviews see e.g.\ \cite{Franosch2013,Metzler_2014,Krapf_2015,Manzo2015RPP,Meroz_2015}).

An important 
class of anomalous diffusion processes are multi-state
dynamics -- dynamics that depend on the instantaneous value of an 
internal state. Examples of multi-state dynamics include
stochastically gated diffusion-controlled reactions
\cite{Szabo_1982,Zwanzig_1992,Gated,Gated2,Emerging_JPA}, intermittent
molecular search processes underlying cellular gene regulation
\cite{Mirny,Koslover_2011,Sheinman_2012}, or the center of mass
diffusion of (un)folding proteins \cite{Yamamoto} or growing polymers
\cite{Baldovin_2019}, the anomalous diffusion of membrane proteins \cite{Maizon2021BiB}, as well as tracer diffusion in heterogeneous
environments \cite{Kaerger,Grebenkov}, and most recently also in
mammalian cells \cite{Weiss2020,Janczura_2021}.

Notwithstanding all progress in the development of analytical tools
\cite{Eli_PhysToday,Franosch2013,Metzler_2014,Ernst_2014,Meroz_2015}
it often remains challenging to conclusively distinguish
between different modes of motion observed in experiments \cite{Woringer_2020}, and in
particular to conclusively identify multi-state transport
\cite{Sikora2017PRE,Akimoto2017PRE,Lanoiselee2017PRE}. To this end, we focus  on scattering
fingerprints of multi-state dynamics, that is, on the intermediate
scattering function (ISF) and dynamic structure 
factor (DSF), observables that are typically monitored in neutron and X-ray
scattering experiments (for applications in the context of biophysical
systems see e.g.\
\cite{Zaccai2000,Kneller2005,Biehl2011,Martel1992PBmB}).
The ISF and the DSF were originally identified as key observables in
scattering experiments by van Hove
\cite{VanHove1954}. While the ISF was initially introduced as an auxiliary
observable in the study of the DSF, more recent methods, such as
neutron spin echo (NSE) spectroscopy
\cite{Mezei2002book,Richter2005,Callaway2017,Liu2017}, as well as
modern (colloidal) particle-tracking techniques, such as
differential dynamic microscopy (DDM)
\cite{Kurzthaler2018,Cerbino2008,Wilson2011} or Fourier imaging correlation
spectroscopy (FICS) \cite{Senning2010,Knowles2000,Kolin2006}, probe
the ISF directly.

Motivated by recent experimental observations (see e.g.\ \cite{Weiss2020,Janczura_2021,Monkenbusch2016})
our aim is to employ the ISF
and DSF
in the analysis of multi-state transport, more precisely, 
of two-state dynamics
. By combining theory and particle-tracking experiments in living cells we consider inert tracer-particle dynamics
as well as systems with internal degrees of freedom.
Our results 
may be relevant for the
interpretation of DDM
experiments on complex particulate systems, as well as neutron and X-ray scattering
scattering, when the underlying dynamics displays two-state transport.  

\section{Theory}
\subsection{Intermediate scattering function and the dynamic structure 
factor}

The fundamental quantities, the ISF and DSF, are calculated from the
so-called van Hove function $G(\vec r,t)$, which is a generalized pair
distribution function \cite{VanHove1954,SimpleLiquidsBook2013}. In the absence of quantum effects 
$G(\vec r,t)$ is the density-density time-correlation
function that may be interpreted as the average number of particles (scatterers) $\beta$ in a region $d\vec r$ around
a point $\vec r+\vec r\,'$ at time $t$ given that there was a particle (scatterer) $\alpha$ at
a point $\vec r\,'$ at time $t = 0$ (whereby the initial condition $\vec r\,'$ is averaged out),
\begin{equation}
  G^{\alpha\beta}(\vec r,t)=\langle\delta(\vec r_\beta(t)-\vec r_\alpha(0)-\vec r)\rangle.
  \label{vanHove-component}
\end{equation}
The brackets denote the average over an
appropriate ensemble
that we will specify below. For the sake
of simplicity we here
focus exclusively on systems of non-interacting molecules, which,
however, may possess internal degrees of freedom. 
Depending on
whether $\alpha=\beta$ or not (i.e.\ whether we consider scattering events at
an individual or two distinct
scatterers) the van Hove function splits into an incoherent part (self-part, $\alpha=\beta$) that probes single-scatterer motions, and a coherent part (sum of  distinct-part, $\alpha\ne\beta$, and self-part) probing collective motions \cite{SimpleLiquidsBook2013}. More precisely, 
\begin{align}
G_\text{inc}(\vec r,t)&=\frac{1}{N}\sum_{\alpha=1}^N G^{\alpha\alpha}(\vec r,t), \label{vanHove-inc}\\
G_\text{coh}(\vec r,t)&=\frac{1}{N}\sum_{\alpha,\beta=1}^N G^{\alpha\beta}(\vec r,t), \label{vanHove-coh}
\end{align}
where the sum runs over all $N$ scattering centers within a molecule
and we will in addition average over an ensemble of
statistically independent trajectories of the molecule (in fact
  over an
  ensemble of many such molecules) that we denote
by $\langle\cdot\rangle$. Depending on the experimental setup
$G_\text{inc}(\vec r,t)$ and $G_\text{coh}(\vec r,t)$ may
\cite{Reat1998} or may not be monitored
separately.

The intermediate scattering function (ISF; mathematically the
characteristic function of displacements\footnote{If moments of the
displacement are finite, the ISF as a function of $q$ contains the complete information about
the statistics of the displacements.}) measured in NSE,
DDM and
FICS is the spatial Fourier transform of $G^{\alpha\beta}(\vec r,t)$,
\begin{equation}
F^{\alpha\beta}(\vec q,t)=\int\rmd^3 r\ G^{\alpha\beta}(\vec r,t)\rme^{-i\vec q\cdot \vec r}=\langle \rme^{-i\vec q\cdot (\vec r_\beta(t)-\vec r_\alpha(0))}\rangle,
\label{ISF-component}
\end{equation}
while the dynamic structure factor (DSF) that is measured in
inelastic/quasielastic scattering experiments is the space-and-time
Fourier transform of $G^{\alpha\beta}(\vec r,t)$, i.e.\
\begin{equation}
    S^{\alpha\beta}(\vec q,\om)=\frac{1}{2\pi}\int_{-\infty}^{\infty}\rmd t\ \rme^{i\om t}F^{\alpha\beta}(\vec q,|t|)=\frac{1}{\pi}\int_{0}^{\infty}\rmd t\ \cos(\om t)F^{\alpha\beta}(\vec q,t),
\label{DSF-component}
\end{equation}
or alternatively defined via the Laplace transform $\widehat{f}(s)\equiv
\int_0^{\infty}{\rm e}^{-st}f(t){\rm d}t$ as $S^{\alpha\beta}(\vec q,\om)=\pi^{-1}{\rm
  Re}[\widehat{F^{\alpha\beta}}(\vec q,s=-i\omega)]$, where ${\rm
  Re}$ denotes the real part.

In scattering experiments the arguments $\vec q$ and $\omega$ correspond to the momentum and
energy transfer 
and the DSF is
proportional to the measured intensity \cite{VanHove1954}. In the next
step we develop results for $F^{\alpha\beta}(\vec q,t)$ and $ S^{\alpha\beta}(\vec
q,\om)$ for systems displaying two-state dynamics. 

\subsection{Two-state Dynamics}\label{twoc}
Assuming two distinct dynamic ``states'' interconverting in continuous time in a
Markovian manner with rates $k_1,k_2>0$, that is, according to the
master equation $\dot{\vec p}(t)={\rm K}\vec p(t)$ with transition rate matrix
\begin{align}
    {\rm K}
    =\begin{bmatrix} -k_1 & k_2 \\ k_1 & -k_2 \end{bmatrix}.
\end{align}
Using 
\begin{align}
    p_1^{\rm eq}&\defas\frac{k_2}{k_1+k_2},\qquad p_2^{\rm eq}\defas\frac{k_1}{k_1+k_2},
\label{p_i}\end{align}
the
jump-propagator is diagonalized as 
\begin{equation}
    \exp({\rm K}t)=\binom{p_1^{\rm eq}}{p_2^{\rm eq}}(1,1)+\rme^{-(k_1+k_2)t}\binom{1}{-1}\left (p_2^{\rm eq},-p_1^{\rm eq}\right ).
\end{equation}

We now assume $N=1$ and drop the indices $\alpha,\beta$; the case
$N>1$ is discussed in \ref{apSingwi}. Let $G_j(\vec r,t)\psi_j(t)$ denote the joint density to observe a displacement $\vec r$ in
a time $t$ in dynamical state $j$ without changing the state, where
$G_j(\vec r,t)$ denotes the van Hove function for the dynamics in a
single state $j$. Its Fourier-Laplace transform will be denoted by  
\begin{equation}
    \widehat{(F\psi)}_j\equiv\widehat{(F\psi)}_j(\vec q,s)\equiv\int {\rm d}^3 r \int_0^{\infty}{\rm
      d}t\,\mathrm{e}^{-st-i\vec q \cdot \vec r}G_j(\vec
    r,t)\psi_j(t).
    \label{FLap}
\end{equation}
Often the change of state erases all memory in the
  sense that the propagation within each sojourn can be assumed to be independent of
  the configuration just before the switch \cite{Singwi1960}, as in the
  case e.g.\ when the dynamics in both states is translation
  invariant. In this case the DSF and ISF of the
  complete two-state dynamics can be obtained
following Ref.~\cite{Singwi1960}
by a direct summation over all possible
realizations of sojourns until time $t$ 
(for the derivation see \ref{apSingwi}).
The Fourier-Laplace transform of the
complete two-state 
propagator, $G(\vec r,t)$, in this case reads
\begin{equation}
  \widehat{F}(\vec q,s)= \frac{p_1 \widehat{(F\psi)}_1\left[1+k_1\widehat{(F\psi)}_2\right]+p_2 \widehat{(F\psi)}_2\left[1+k_2\widehat{(F\psi)}_1\right]}{1-k_1k_2\widehat{(F\psi)}_1\widehat{(F\psi)}_2},
\label{singwi series}
\end{equation}
where $p_{1,2}$ denote the initial occupation of states. The DSF
henceforth follows
immediately using
\begin{equation}
  S(\vec q,\om)=\frac{1}{\pi}\Re[\widehat{F}(\vec q,-i\om)].
\label{DSF}  
\end{equation}  
The problem of determining the ISF and DSF for two-state dynamics
with ``complete memory erasure'' upon each change of state
thus boils down to determining $\widehat{(F\psi)}_{1,2}(\vec q,s)$ in
Eq.~(\ref{FLap}) for the specific example under consideration. In case
there is no complete  memory erasure (see protein (un)folding example
Sec.~\ref{unfolding}) one must solve the specific model at hand using
a dedicated method or resort to approximations.
In the following we address a set of insightful and 
experimentally relevant scenarios of two-state diffusion.

\section{Results}
\subsection{Two-state diffusion}\label{twoch}

In the first example we assume that the molecule/tracer particle undergoes free Brownian motion 
with a diffusion coefficient that randomly switches between values
$D_1$ and $D_2$ in a Markovian fashion with rates $k_1$ and
$k_2$, implying the distribution $\psi_j(t)=\mathrm{e}^{-k_j t}$ of sojourn times in the respective states. This situation arises in the context of tracer diffusion in heterogeneous media (see
e.g.\ \cite{Kaerger,Grebenkov,Emerging_JPA}). Note that assuming
    sojourn times to be independent of the dynamics within the
    individual states is not always justified. In particular, we
    here assume an ``annealed heterogeneity'' (see e.g. \cite{Alexei}). This may occur when the
    medium is translationally invariant such as in the case of a state change
    triggered by the reversible binding to a homogeneously distributed
    mobile species. In general this does not apply to quenched
    heterogeneous media, except in cases where a system is
    observed under equilibrated conditions \cite{Postnikov}.

The Fourier-Laplace
transform of $G_j(\vec r,t)\psi_j(t)$ defined in
Eq.~(\ref{FLap}) reads
\begin{equation}
  \widehat{(F\psi)}_j(\vec q,s)=\widehat{(F\psi)}_j(q,s)=
  \frac{1}{s+q^2 D_j+k_j},
\label{two}
\end{equation}
where $q^2\equiv \vec q\cdot \vec q$. Plugging Eq.~(\ref{two}) into
Eq.~(\ref{singwi series}) the Fourier-Laplace image of the
complete propagator for two-state diffusion can be written as
\begin{equation}
  \widehat{F}(q,s)=
  \frac{s+q^2(p_1D_2+p_2D_1)+k_1+k_2}{(s+q^2 D_1+k_1)(s+q^2 D_2+k_2)-k_1k_2},
\label{singwi series2}
\end{equation}
with a general initial condition $p_{1,2}$.
Defining 
\begin{equation}
        \mu_{\pm}(q)=\frac{q^2(D_1+D_2)+k_1+k_2}{2}\pm\sqrt{\frac{[q^2(D_1-D_2)+k_1-k_2]^2}{4}+k_1k_2}\,,\,
        \label{mu}
\end{equation}
introducing the auxiliary function
\begin{align}
    \varphi(q)&\defas\frac{q^2(p_1D_2+p_2D_1)+k_1+k_2-\mu_-(q)}{\mu_+(q)-\mu_-(q)},
\end{align}
and inverting into the time domain we finally obtain
\begin{align}
    F(q,t)&=[1-\varphi(q)]{\rm e}^{-\mu_+(q)t}+\varphi(q){\rm e}^{-\mu_-(q)t},\label{diffusions-ISF}\\
    S(q,\om)&=\frac{1}{\pi}\Re[\widehat{F}(q,-i\om)]=\frac{1}{\pi}\frac{[1-\varphi(q)]\mu_+(q)}{\om^2+\mu_+^2(q)}+\frac{1}{\pi}\frac{\varphi(q)\mu_-(q)}{\om^2+\mu_-^2(q)}.\label{diffusions-DSF}
\end{align}
In contrast, in the case when the two states co-evolve as a ``frozen
mixture'' --- an ensemble consisting of two types of molecules that do not
interconvert between each other --- 
with the state occupations $p_{1,2}$ we have $F_{\rm
  mix}(q,t)=p_1\mathrm{e}^{-q^2D_1t}+p_2\mathrm{e}^{-q^2D_2t}$. This
frozen mixture will throughout be denoted by the subscript
``mix'', and is the limit of Eq.~\eqref{diffusions-ISF} in the
  case of slow switching rates $k_1+k_2\ll q^2D_{1,2}$, see \ref{apDiffusions}.

It is also instructive to inspect the fast-switching limit, $k_1+k_2\gg q^2D_{1,2}$, where we find (for details see \ref{apDiffusions}) $\mu_+(q)\simeq q^2D_{\rm eff}\equiv p^{\rm eq}_1D_1+p^{\rm eq}_2D_2$ and $\mu_-(q)\simeq k_1+k_2\gg q^2D_{1,2}$, which implies simple diffusion with an effective diffusion coefficient in
agreement with intuition.

Our aim here is to demonstrate that the analysis allows to distinguish between single-state dynamics, two-state switching diffusion and the
 ``frozen mixture'' with the same stationary probabilities as long as the
switching rates are not too fast or too slow, which would give rise to
the aforementioned (fast-switching) effective diffusion and ``frozen
mixture'', respectively. 
\begin{figure} \centering
\includegraphics[scale=.8]{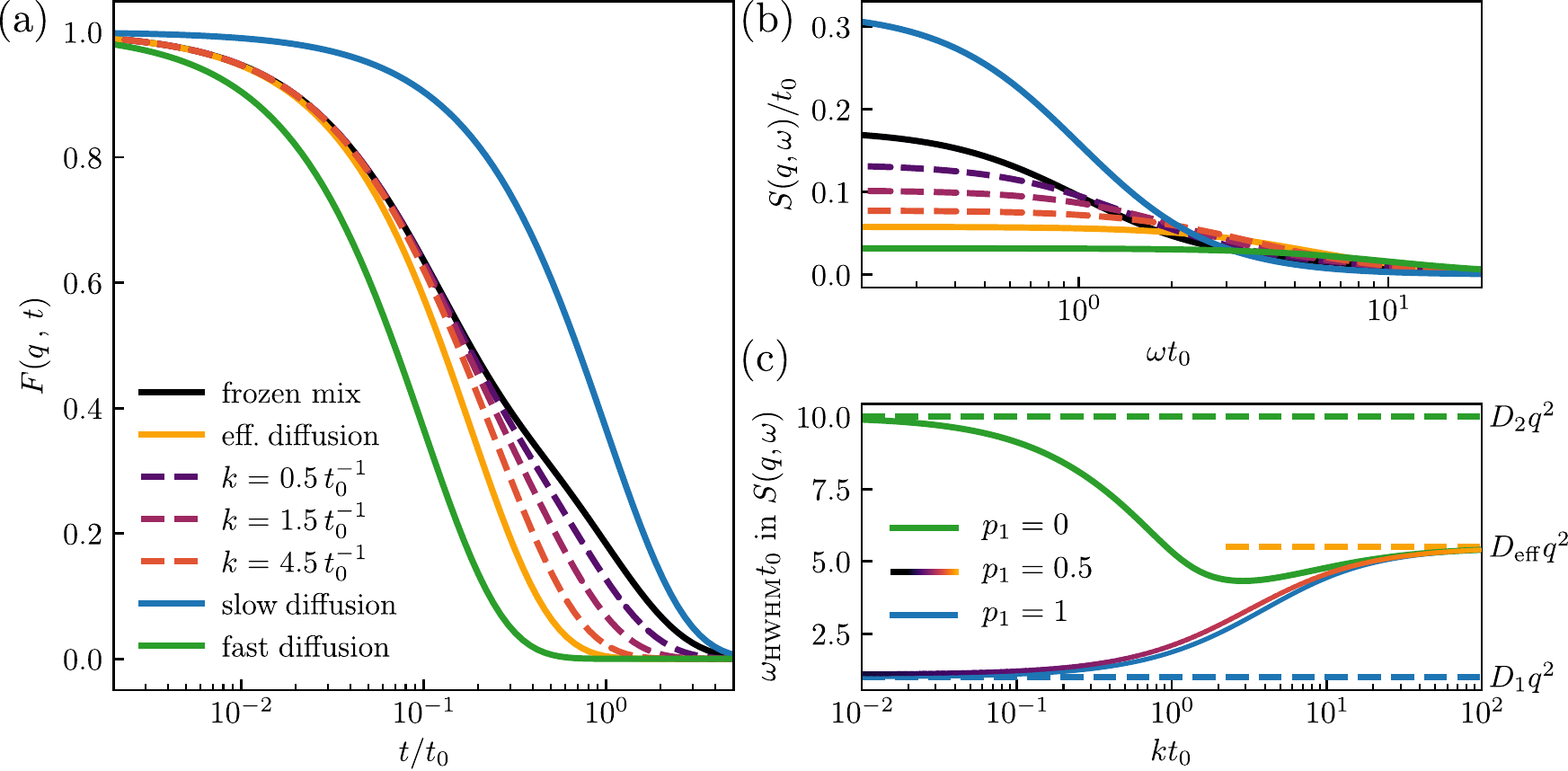}
\caption{\textbf{(a)} ISF $F(q,t)$ for free diffusion
  switching between diffusion constants $D_1$ and $D_2=10D_1$ with switching
  rates $k_1=k_2=k$ evolving from an initial equilibrium occupation
  $p^{\rm eq}_1=p^{\rm eq}_2=0.5$ and time $t$ measured in units of $t_0=1/D_1q^2$. The ISF for different switching rates are shown along with
  the limits of the frozen mix ($k\to 0$, black curve) and the
effective diffusion ($k\to\infty$, yellow curve). \textbf{(b)}
Dynamical structure factor corresponding to the ISF in
(a). \textbf{(c)} Half width at half maximum (HWHM) of the DSF from
(b) are shown along with the HWHMs of the blue, green and yellow
curves in (b) (dashed lines). To compare with a non-equilibrium
initial occupation, the HWHM for initial occupation $p_1=0$ and
$p_1=1$ are also shown (green and blue lines, respectively).}
\label{fg1}
\end{figure}

For such moderate switching rates the ISF shown in
Fig.~\ref{fg1}a and the DSF in Fig.~\ref{fg1}b indeed reveal distinctive
characteristics of two-state diffusion (dashed lines) when compared to the ``frozen
mixture'' (black full line) as well as the two individual diffusive
states (blue and green full lines, respectively). To reduce these differences to a
single observable we
inspect the half width at half the maximum (HWHM) of the
quasielastic peak as a function of the switching rate (that for
simplicity we set to be symmetric, $k_1=k_2\equiv k$) for different initial
conditions. The HWHM is
determined from Eq.~\eqref{diffusions-DSF} by setting
$S(q,\omega_{\rm HWHM})=S(q,0)/2$ and solving the corresponding bi-quadratic equation
for $\omega_{\rm HWHM}$.
The results for $\omega_{\rm HWHM}$ are shown in Fig.~\ref{fg1}c and
together with Fig.~\ref{fg1}a-b
confirm that the DSF and ISF can
reliably distinguish between a ``non-communicating'', frozen superposition of
dynamic modes and two-state diffusion as long as the switching rates
are not too fast or too slow.

Note that for an isotropic
  Gaussian process we have $F(q,t)={\rm e}^{-q^2\langle[\vec r(t)-\vec
    r(0)]^2\rangle/2d}$ and hence ISF contains exactly the same
information as the mean squared displacement (MSD) $\langle[\vec r(t)-\vec
r(0)]^2\rangle$. For the individual states as well as for effective diffusion
the dynamics is Gaussian. However, two-state dynamics is typically
    non-Gaussian \cite{Weiss2020} and the ISF provides more
    information about the dynamics. Note, moreover, that the MSD of
    the two-state process reflects (normal) diffusion
    with effective diffusivity $D_{\rm eff}$ for all switching rates and thus
    alone cannot  reveal underlying two-state dynamics. In the next example we turn to the
switching between two subdiffusive states.

\subsection{Two-state Fractional Brownian Motion}

Tracer transport in complex, dynamic and/or heterogeneous media such
as living cells is often found to be anomalous
\cite{Eli_PhysToday,Franosch2013,Weiss2014IRCMB,Metzler_2014}. Particularly interesting are
situations with stochastically interconverting anomalous diffusion
processes such as the two-state anti-persistent (i.e.\ subdiffusive)
fractional Brownian motion observed in recent experiments on mammalian
cells \cite{Weiss2020,Janczura_2021}. 
To be more precise, a systematic analysis of the motion of quantum dots immersed in the
cytoplasm of living mammalian cells revealed
two-state subdiffusive
fractional Brownian motion with exponent $\alpha<1$ but with a
stochastically switching anomalous diffusion coefficient $C_{1,2}$
\cite{Weiss2020}. Here we determine the DSF and ISF for the process as
complementary observables and apply them to the analysis of data
obtained in \cite{Weiss2020} focusing on latrunculin-treated cells.  Untreated cells had been seen to feature the same dynamics \cite{Weiss2020}. Note that here a Markov switching between the
two states coexists with non-Markovian subdiffusion in each of the
two states, respectively, with a mean squared displacement $\langle [\vec r(t)-\vec
r(0)]^2\rangle_j=2dC_jt^{\alpha_j}$ where $d$ (here equal two) is the dimensionality of the
system, and $\langle\cdot\rangle_j$ denotes the
average over an ensemble of fractional Brownian motions (FBMs) with exponent
$\alpha_j$ and
coefficient $C_j$. 
Thus, there is an inherent ambiguity in how to
treat memory upon each jump. Here, in contrast to the original
publication \cite{Weiss2020}, we assume for convenience that the memory in $\vec r(t)$
is erased upon each change of state, which allows us to use the
result in Eq.~(\ref{singwi series}). Notably, this choice does not affect
the inferred two-state FBM, which further substantiates
the robustness of the proposed model.

We first treat the problem theoretically. The $d$-dimensional
fractional Brownian motion is an isotropic and translationally invariant Gaussian process. Thus
the ISF of FBM is the characteristic function of a Gaussian process
$F_j(q,t)=\exp(-q^2 \langle [\vec r(t)-\vec r(0)]^2\rangle_j/2d)$. 
Within a sojourn in 
state $j=1,2$ the characteristic function of the displacement $\vec r$ in
a time $t$ without changing the state is $F_j(q,t)\psi_j(t)=\exp(-[q^2 C_j t^{\alpha_j}+k_j]t)$. 
Its Laplace transform $\widehat{(F\psi)}_j(q,s)$ may be written in
terms of Meijer G-functions in \ref{apFBM} (see Eq.~\eqref{Meijer}). Plugging the result for $\widehat{(F\psi)}_j(q,s)$ into Eq.~(\ref{singwi series}) yields the Laplace-transformed ISF, which we invert
numerically into the time domain. The DSF is in turn obtained fully analytically by
setting $s=-i\om$ and taking the real part as in Eq.~\eqref{DSF}.

\begin{figure} \centering
\includegraphics[scale=0.8]{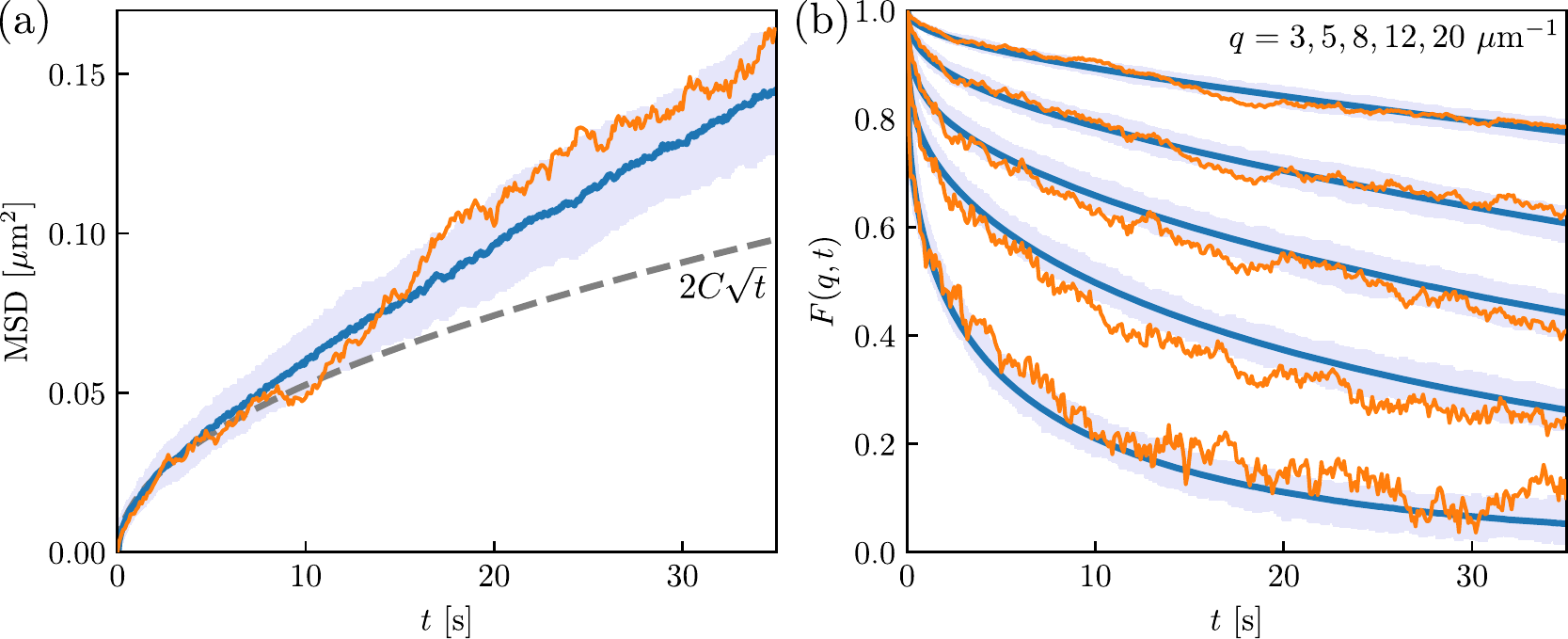}
\caption{\textbf{(a)} Mean-squared displacement (MSD) derived from
  the experiment
  (orange curve) and MSD simulated with fit parameters from (b) (blue
  curve with blue shaded standard deviation). The grey dashed line
  shows a simple subdiffusive $\sqrt{t}$-fit to the first 10 seconds,
  yielding $C=0.0083\ \mu{\rm m}^2/\sqrt{\rm s}$. \textbf{(b)} ISF $F(q,t)$ for (from top to bottom)
  $q=3,5,8,12,20\ \mu{\rm m}^{-1}$ from experiment (orange curve) and
  theory (blue curve) with standard deviation (blue shaded) obtained from
  simulations (experimental and simulation details are given in \ref{apSim});
  the fitting procedure and parameters are given in the
  text; the range of $q$ values was selected to capture the
  spatio-temporal scales in the relaxation observed experimentally,
  i.e.\ such that $F(q,t)$ spans the full range of values.}
\label{fg2}
\end{figure}

The model contains six parameters $C_1,C_2,k_1,k_2,\alpha_1,\alpha_2$
to be determined. To minimize overfitting we fix three
parameters by considering the short-time behavior of the MSD, see
dashed line in  Fig.~\ref{fg2}a. We see that the short time MSD
(i.e.\ prior to any change of state)  is
well described by simple subdiffusion with $\alpha=0.5$, and we thus
fit at the first $10$ seconds yielding $C_{\rm short}=0.0083\ \mu{\rm
  m}^2/\sqrt{\rm s}$.  To reduce the number of free parameters, we
constrain ourselves to exactly reproduce this short-time behavior at
times sufficiently short to not be influenced by transitions, i.e.\ we
fix $\alpha_1=\alpha_2=0.5$ and $p_1^{\rm eq}C_1+p_2^{\rm eq}C_2=C_{\rm short}$ (recall
$p_1^{\rm eq}=k_2/(k_1+k_2)$ from Eq.~\eqref{p_i}). This leaves three free parameters which we
determine by a least-squares fit simultaneously at $q=3$ $\mu{\rm m}^{-1}$ and
$q=7$ $\mu{\rm m}^{-1}$. The fitting yields
$C_1=0.002\ \mu{\rm
  m}^2/\sqrt{\rm s},\ C_2=0.052\ \mu{\rm
  m}^2/\sqrt{\rm s},\ k_1=0.009\ {\rm s}^{-1},\ k_2=0.063\ {\rm s}^{-1}$.

For the case $\alpha_j=1/2$ the Fourier-Laplace transform
of the joint density to observe a displacement $\vec r$ in
a time $t$ without changing the state
reads (see \ref{apFBM})
\begin{align}
    \widehat{(F\psi)}_j(q,s)=\frac{1}{s+k_j}-\frac{\sqrt{\pi}q^2 C_j}{2(s+k_j)^{3/2}}\text{erfcx}\left(\frac{q^2 C_j}{2\sqrt{s+k_j}}\right),
\label{fbm_joint}
\end{align}
where $\text{erfcx}(x)\equiv\exp(x^2)(1-\text{erf}(x))$ and erf denotes the
error function. Via Eq.~\eqref{singwi series} and numerical Laplace inversion this yields the two-state ISF. The theoretical ISF (blue lines) alongside the statistical uncertainty expected from a set of 400 realizations (shaded area)
and the ISF determined directly from the experimental trajectories is
shown in Fig.~\ref{fg2}b for several values of $q$. The 
comparison between the corresponding mean squared displacements is
shown in Fig.~\ref{fg2}a. The theoretical fit and experimental results display
a good agreement. In particular, the analysis of the ISF not only
confirms the findings in Ref.~\cite{Weiss2020} but also demonstrates the
two-state FBM to be an appropriate model for the observed dynamics over a
broad range of spatial and temporal scales. Notably, this
spatio-temporally resolved information is inherently coarse-grained
out in the analysis of the mean squared displacement. Recall that because the
process is not Gaussian \cite{Weiss2020} the ISF and DSF, but not the MSD,
contain the full information about the dynamics. The scattering
fingerprints proposed here are thus well suited for a deeper analysis
of particle-tracking experiments displaying multi-state anomalous diffusion.

\subsection{Reversible dimerization with an internal mode}\label{dimerization}
\begin{figure} \centering
\includegraphics[scale=0.8]{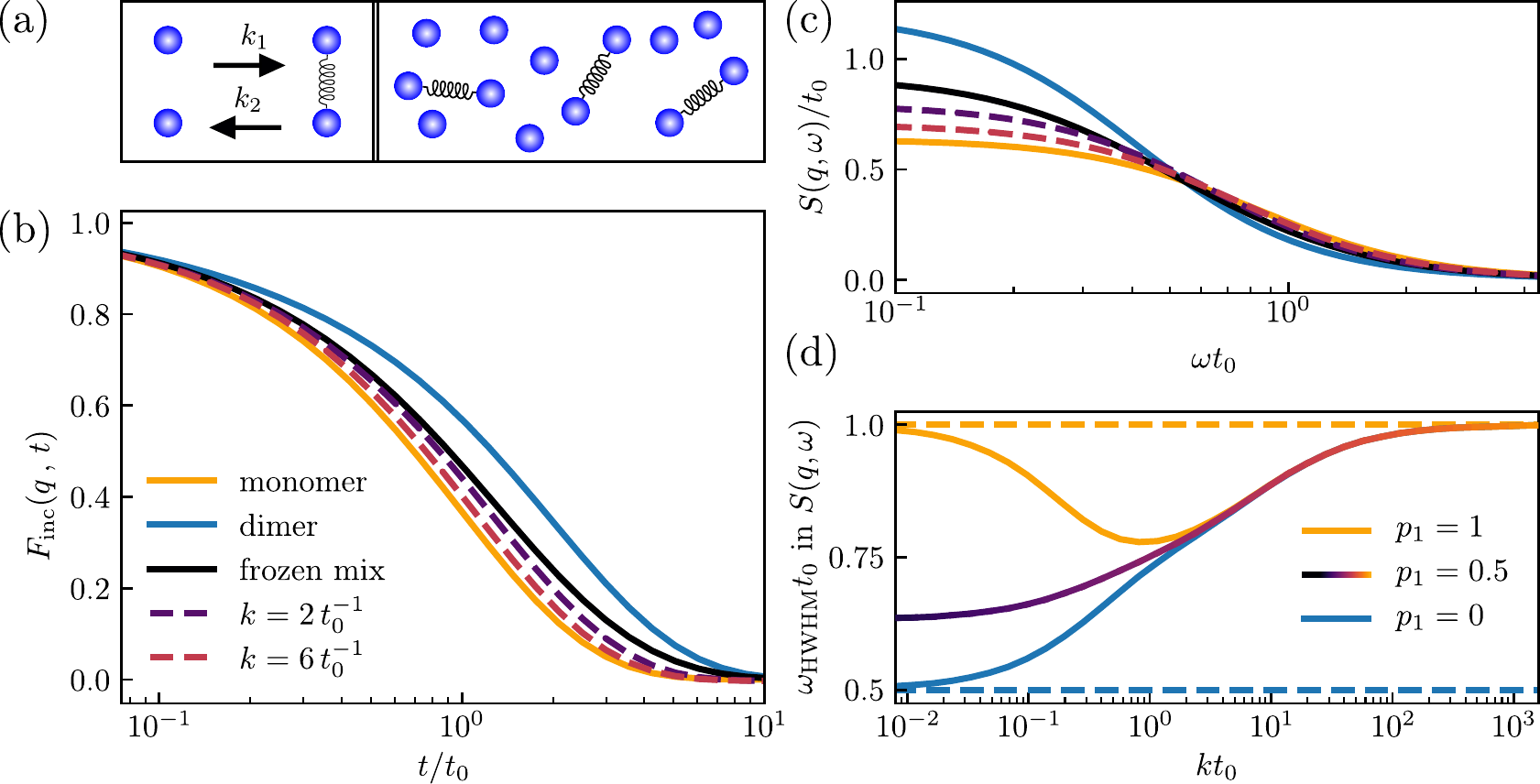}
\caption{\textbf{(a)} Schematic of the dimerization process. \textbf{(b)} Incoherent ISF $F_{\rm inc}(q,t)$ for reversible dimerization of a two-bead (Rouse) dimer with $k=k_1=k_2$ evolving from an equilibrium occupation $p_1=p_2=0.5$ and time $t$ measured in units of $t_0=1/Dq^2$. The limit for $k\to 0$ is the frozen mix (black curve) and for $k\to\infty$ the ISF approaches the ISF of a monomer without dimerization (yellow curve) since the unbinding occurs before the internal mode enters the dynamics. For the explicit derivation of this limit see \ref{apDimer}. \textbf{(c)} DSFs corresponding to (a). \textbf{(d)} HWHMs of the DSF from (c) are shown along with the HWHMs of the blue and yellow curves in (c) (dashed lines). The HWHMs for initial occupation $p_1=0$ and $p_1=1$ are also shown (blue and yellow lines).}
\label{fg3}
\end{figure}

So far we have only considered two-state dynamics without any
internal degrees of freedom. To go beyond this limitation, we consider diffusion in
the presence of reversible dimerization (see Fig.~\ref{fg3}a) captured in the mean field
limit -- two particles are assumed to associate
with an effective rate that is independent of the particles' instantaneous
position in the spirit of Smoluchowski \cite{Smoluchowski1916,Collins1949JCS}. To be concrete, we consider non-interacting
particles diffusing freely with a diffusion coefficient $D_1=D$ and
forming a dimer with a center-of-mass diffusion coefficient
$D_2=D/2$ and an internal harmonic vibrational relaxation mode with rate $a$
(i.e.\ the internal coordinate -- the distance between the associated
particles -- evolves as an Ornstein-Uhlenbeck process
\cite{Ornstein1930} or Rouse model
with $N=2$). Note that since $N>1$ in contrast to the previous examples we here need to
distinguish between coherent (i.e.\ Eq.~(\ref{vanHove-coh})) and incoherent (i.e.\ Eq.~(\ref{vanHove-inc})) contributions. We focus on 
the incoherent part (i.e.\ we set $\alpha=\beta$ in
Eq.~\eqref{vanHove-component}) and note that the beads are equal and thus the beads $\alpha=1$ and $\alpha=2$ give an
identical scattering contribution.

The association and dissociation rates are denoted by $k_1$
and $k_2$ respectively. Since the internal dynamics does not depend on
the absolute position in space and each change of state
erases all memory we can employ the result in Sec.~\ref{twoc}. In
particular, this assumption neglects a possibly enhanced probability for
re-association of the same pair of molecules immediately upon
dissociation, which is expected to be a good approximation in well-mixed (i.e.~``stirred'') systems
and/or at high monomer concentrations. The opposite case, when
re-association is more likely, introduces at least one more
time scale in to the kinetics of association, thus rendering the
two-state switching process non-Markovian. For the sake of simplicity
we stick to the Markovian scenario.

The
Fourier-Laplace transforms of the joint density to observe a displacement $\vec r$ in
a time $t$ in the respective state 1 prior to switching,
$\widehat{(F\psi)}_1(q,s)$, is given by Eq.~(\ref{two}) with
$D_1=D$. Conversely, introducing $s^\prime=s+q^2D/2+k_2$ (note that
$s'=s'(q)$ depends on $q$) we can write $\widehat{(F\psi)}_2(q,s)$  as
(for a derivation see \ref{apDimer})
\begin{eqnarray}
\widehat{(F\psi)}_2(q,s')&=&\frac{1}{s'}\rme^{-q^2D/2a}\left(q^4 D^2/4a^2\right)^{s'/a}\!M\left(\frac{s'}{a},\frac{s'+1}{a},\frac{q^2D}{2a}\right)\nonumber\\&=&\rme^{-q^2
  D/2a}\sum_{n=0}^\infty\,\frac{(q^2D/2a)^{n}}{n!}\frac{1}{s'+na},
\label{dimer}
\end{eqnarray}
where 
$M(a,b,z)$ denotes Kummer's confluent hypergeometric function
\cite{Bateman}. Using $\widehat{(F\psi)}_1(\vec q,s)$ and
$\widehat{(F\psi)}_2(\vec q,s')$ in Eq.~(\ref{singwi series}) we find the
Laplace image of the incoherent ISF and, via Eq.~(\ref{DSF}), also the incoherent DSF, where
we recall that here the coherent and incoherent contribution do not
coincide. The ISF may in principle be obtained by analytical inversion
of the Laplace transform. For the sake of simplicity we here perform
the inversion numerically with the fixed Talbot method.

The results for the incoherent ISF for the two-state dynamics
with $k_1=k_2=k$ alongside the
frozen mixture and dynamics in the individual pure states,
all evolving from an equilibrium initial condition, 
are shown in Fig.~\ref{fg3}b.  The corresponding DSF is depicted in
Fig.~\ref{fg3}c and the comparison is further clarified by means of
the HWHM shown in Fig.~\ref{fg3}d. 
Altogether, one can distinguish
between the different situations as long as $k$ is not too small or
too large. These results demonstrate that the scattering fingerprints proposed here
may be a valuable tool for analyzing two-state dynamics also in the
presence of internal motions. 

\subsection{Two-state dynamics with internal structural motions -- protein (un)folding}\label{unfolding}
\begin{figure}\centering
\includegraphics[scale=0.9]{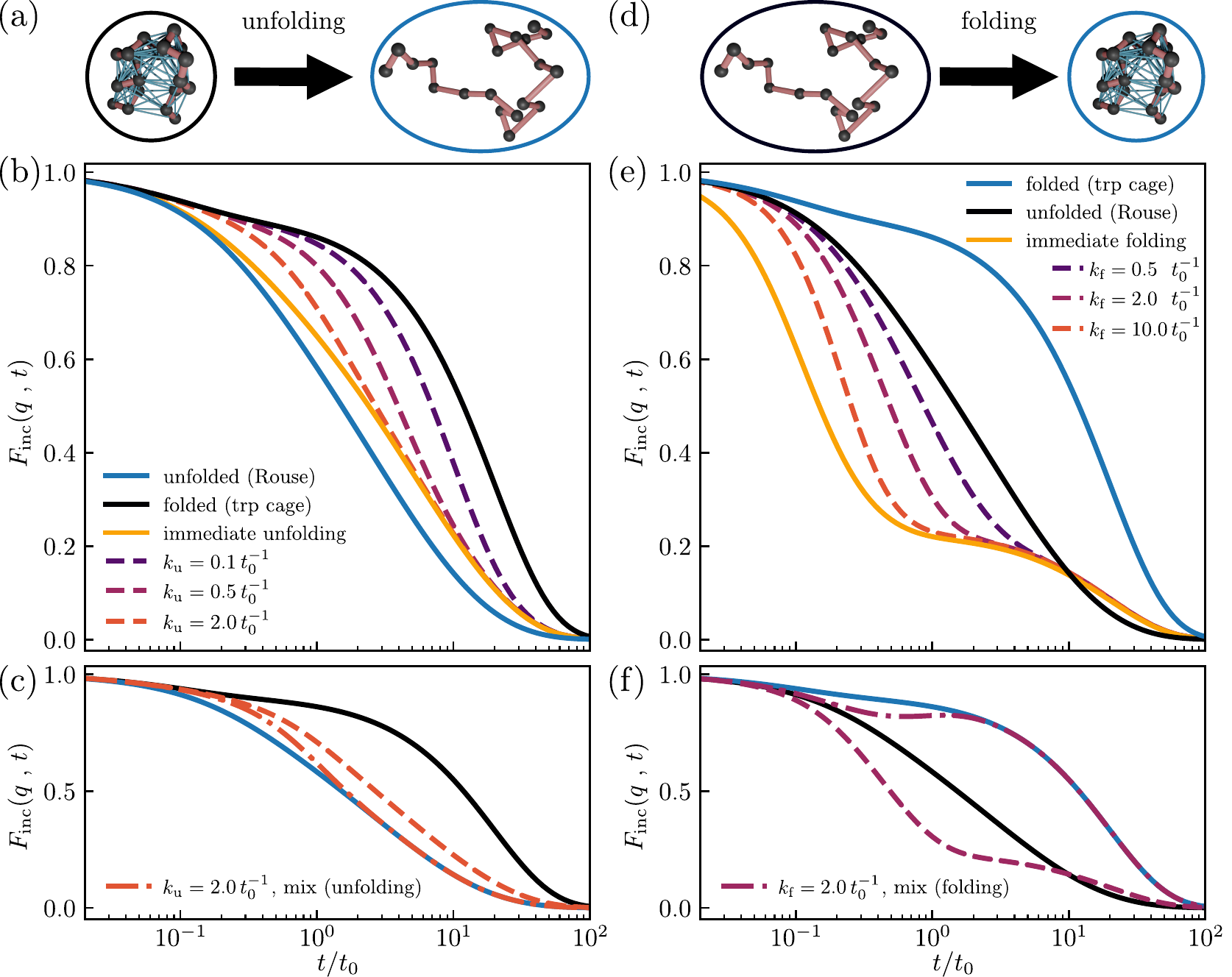}
\caption{\textbf{(a)} Schematic of the
  Gaussian network model for the
  Trp-Cage Miniprotein Construct TC5b (PDB: 1L2Y) and its unfolding to
  a Rouse chain ($N=20$). \textbf{(b)} Incoherent ISF $F_{\rm
    inc}(q,t)$ for the (non-reversible) unfolding
  process with time measured in units of $t_0=1/Dq^2$ and a ratio of spring and friction constant of $k/\xi=t_0^{-1}$. The dashed lines denotes various unfolding rates $k_u$, and
  the yellow curve, the "immediate unfolding" limit, is obtained
  by setting $\tau=0$ in Eq.~\eqref{all terms}. Also depicted are the $F_{\rm
    inc}(q,t)$ of the respective relaxation processes of the folded
  protein (black line) and the Rouse chain (blue line). Time is
  expressed in standard dimensionless Rouse units (see
  e.g.\ \cite{Lapolla_2021}). \textbf{(c)} Comparison of the
  unfolding process (dashed line)  with the time-dependent "mix"
  (dash-dotted line) which corresponds to the superposition of the ISF of
  the Rouse chain and the protein weighted by the respective
  time-dependent occupation of the state. \textbf{(d-f)} As in (a-c)
  but for the converse folding process.}
\label{fg4}
\end{figure}

A generalization of the preceding results to two-state dynamics in
the presence of some general internal structural relaxation, such as,
for example, in the case of cyclization of polymers, and the
reversible (un)folding or association of proteins, to name but a few,
is much more difficult. In particular, when the change of state does
not erase 
all memory such that the propagation within each sojourn depends on
  the internal configuration just before the switch, the results of
  Sec.~\ref{twoc} cannot be applied. Such a situation naturally
  arises in systems with internal (e.g.~conformational) dynamics in
  absence of a complete time-scale separation, i.e.~when at the time
  of the jump the internal
  degrees of freedom have not yet completely relaxed from their initial
  condition or the condition immediately after the preceding change of
  channel, respectively. 
If one is to nevertheless apply a version of the theory developed here,
additional simplifying assumptions are required. In particular, we
must assume that the switching rate is independent of the instantaneous
structural state in either dynamical state.
Even in this case the switching between
states at any instance occurs from a distribution of
structures that set the initial conditions for the relaxation
following the jump. Further, as long as the dynamical states have substantially
different equilibrium structures, the ISF is expected to depend qualitatively on $q$ (i.e.\ large and intermediate scale motions are expected to
differ in contrast to individual bond vibrations at large $q$). 
In full generality this corresponds to a highly non-trivial
problem. To render the general problem analytically tractable one may, for example, inspect the limit of slow jumps that
corresponds to the switching between two equilibrium populations of
internal structures. However, this limit is rather uninteresting as
the state switching is too slow to mix the dynamics of the
individual states, i.e.\ deviations form a ``frozen mix'' are small.
However, it is easy to instead look at the
 two-state dynamics with unidirectional/irreversible jumps evolving from a pure
state (i.e.\ from either of the states). A neat and
experimentally interesting problem is the denaturant-driven
unfolding \cite{Unfolding,Unfolding_2} or folding \cite{Folding} of
proteins. To this end we analyze the irreversible unfolding/folding of
the small 20-residue Trp-Cage protein construct TC5b into/from a Rouse
chain ($N=20$). That is, we consider a sample initially
prepared in the folded state that unfolds with rate $k_{\rm u}$ (Fig.~\ref{fg4}a), and
an initially unfolded (Rouse chain) state folding with a rate $k_{\rm
  f}$, respectively (Fig.~\ref{fg4}d).

A Gaussian elastic network model \cite{Gaussian,Lapolla_2021} of TCPb in the
folded state was constructed from the Protein Data Bank (PDB entry:
1L2Y) using the ProDy software \cite{prody}, yielding a 
Kirchoff matrix ${\rm A^f}$ with dimensional non-zero entries that are
integer multiples of  $k/\xi$ where
$k$ and $\xi$ are the spring and friction constant, respectively.  
Conversely, Rouse chain
dynamics with connectivity matrix ${\rm A^u}$ \cite{Rouse} were
assumed in the unfolded state. ${\rm A^f}$ and ${\rm A^u}$ are given explicitly in \ref{apENM}. 
  For simplicity, in both cases the diffusion matrix is
supposed to be equal and diagonal (isotropic) with diffusion coefficient $D=k_BT/\xi$
for each individual bead. A schematic of the bead-spring model is shown in
Fig.~\ref{fg4}a and d.

Introducing the 3N-dimensional vector of beads' positions
$\mathbf{R}\equiv\{\vec{R}^\alpha\}_{\alpha=1,\ldots,N}$ the equations of motion in each
state follow a 3N-dimensional Ornstein-Uhlenbeck process with positive semi-definite drift
matrix $\A^{\rm f/u}$, i.e.\ $\mathrm{d}\mathbf{R}_t=-\A^{\rm f/u}\mathbf{R}_t{\rm
  d}t+\sqrt{2D}{\rm d}\mathbf{W}_t$ in the respective state ${\rm
  f}$ and ${\rm u}$, where $\mathbf{W}$ denotes a $3N$
dimensional vector whose entries are statistically independent Wiener
processes $W_t$. This process describes the overdamped thermal motion of beads connected by Hookean springs with zero rest length. The analysis is most easily carried out in the respective normal
coordinates $\mathbf{X}^{\rm f/u}=(\Q^{\rm f/u})^T\mathbf{R}$ with
$(\Q^{\rm f/u})^T\A^{\rm f/u}\Q^{\rm f/u}={\rm diag}(a_\alpha^{\rm
  f/u})_{\alpha=1,\ldots,N}$. For details see \ref{apENM}. 
We choose $\alpha=1$ to represent the centre-of-mass mode. The center-of-mass
diffusion coefficient is given by $D_{\rm COM}=D/N$ and within this model does not
depend on the conformation of the protein.
For all but the center-of-mass mode the equilibrium
probability density function is Gaussian with zero mean and
variance $\variance^\alpha_{\rm f/u}$.

To illustrate the theory it suffices to consider the folding-process
alone. Let us consider the propagation in the unfolded state {\rm u} evolving
from the Rouse chain equilibrium distribution. 
We are interested in the $\alpha\beta$-component of the ISF of the
folding process, $F^{\alpha\beta}(\vec q,t)$, assuming no or a single jump occurring at time
$\tau$ with $0<\tau<t$, where $\tau$ is exponentially distributed with
rate $k$. Let $F^{\alpha\beta}_{\rm u}(\vec
q,t)$ be the ISF of the pure unfolded state (Rouse) dynamics. As we
have excluded jumps back to the folded state it is obvious that
$F^{\alpha\beta}(\vec q,t)$ corresponds to
$F^{\alpha\beta}_{\rm u}(\vec q,t)$ weighted by the probability $\rme^{-kt}$ that no jump has occurred until time $t$, and a contribution of
the folding. The latter corresponds to the dynamics in the folded
state starting with a jump at the time $\tau$ and evolving from the equilibrium
distribution of the unfolded (Rouse chain) state, $F^{\alpha\beta}_\utof(\vec q,t;\tau)$, averaged over the
exponentially distributed jumping times \footnote{It is
straightforward to include non-exponential jumping-time distributions.}, i.e.
\begin{align}
F^{\alpha\beta}(\vec q,t)&=\rme^{-kt}F^{\alpha\beta}_{\rm u}(\vec
q,t)+\int_0^t\rmd\tau k\rme^{-k\tau}F^{\alpha\beta}_{\utof}(\vec q,t;\tau).
\label{F with single jump ansatz}
\end{align}
Since the Ornstein-Uhlenbeck process with a Gaussian initial condition is a
Gaussian process, $F^{\alpha\beta}_{\rm u}(\vec q,t)$ and
$F^{\alpha\beta}_\utof(\vec q,t;\tau)$ are characteristic functions
(see e.g.\ Eq.~\eqref{ISF-component}) of
zero mean Gaussian displacements and thus
simply equal to
${\rm exp}{[-q^2\langle(\vec R^\alpha_t-\vec R^\beta_0)^2\rangle_{{\rm u}/\utauf}/6]}$, see \ref{apENM}.
Via normal mode analysis (see, e.g.\ \cite{Edwards}) we show in \ref{apENM} that the displacement
vectors $ \langle(\vec R^\alpha_t-\vec R^\beta_0)^2\rangle_{{\rm u}/\utauf}$ obey
\begin{align}
\left\langle\left(\vec R^\alpha_t-\vec R^\beta_0\right)^2\right\rangle_{\rm u}
&=6D_{\rm COM}t+\sum_{\gamma=2}^N\variance_{\rm u}^\gamma\left[(Q_{\alpha\gamma}^{\rm u})^2+(Q_{\beta\gamma}^{\rm u})^2-2Q^{\rm u}_{\alpha\gamma}Q^{\rm u}_{\beta\gamma}\rme^{-a_\gamma^{\rm u}t}\right],
\label{displacement single process} 
\end{align}
when the Rouse polymer does not fold until $t$, whereas in the case
that folding occurs at a fixed time $\tau$ we find, introducing the
notation $\Q^{{\rm u}\to\rm f}=(\Q^{\rm f})^T\Q^{\rm u}$
(s.t.\ analogous to
$\f R_t=\Q^{\rm f/u}\f X_t^{\rm f/u}$ we have $\f
X_t^{\rm f}=\Q^{{\rm u}\to\rm f}\f X_t^{\rm u}$),
\begin{align}
&\left\langle\left(\vec R^\alpha_t -\vec R^\beta_0\right)^2\right\rangle_\utauf
=6D_{\rm COM}t+\sum_{\gamma=2}^N(Q^{\rm u}_{\beta\gamma})^2\variance_{\rm u}^{\gamma}+\sum_{\gamma=2}^N(Q^{\rm f}_{\alpha\gamma})^2\variance_{\rm f}^{\gamma}\left[1-\rme^{-2a_\gamma^{\rm f}(t-\tau)}\right]\nonumber\\
&+\sum_{\omega=2}^N\variance_{\rm u}^{\omega}\left[\sum_{\gamma=2}^N Q^{\rm f}_{\alpha\gamma}Q^{{\rm u}\to\rm f}_{\gamma\omega}\rme^{-a^{\rm f}_\gamma(t-\tau)}\right]^2-2\sum_{\omega,\gamma=2}^NQ^{\rm f}_{\alpha\omega}Q^{{\rm u}\to\rm f}_{\omega\gamma}Q^{\rm u}_{\beta\gamma}\variance_{\rm u}^{\gamma}\rme^{-a^{\rm f}_\omega(t-\tau)-a^{\rm u}_\gamma\tau}.
\label{all terms} 
\end{align}
The converse unfolding process is obtained by
interchanging ${\rm u}\leftrightarrow\rm f$. Plugging
Eqs.~\eqref{displacement single process}-\eqref{all terms} into Eq.~\eqref{F with single jump ansatz} and performing a numerical integration over $\tau$ yields the result for $F^{\alpha\beta}(\vec q,t)$. Summations over $\alpha,\beta$ deliver the (in)coherent ISF, see Eqs.~\eqref{vanHove-inc}-\eqref{vanHove-coh}.

In Figs.~\ref{fg4}b-c and Fig.~\ref{fg4}e-f  we show, respectively, the incoherent
part of the ISF for various unfolding and folding rates.
The limit of large (un)folding rates is given by setting $\tau=0$ in
the Gaussian ISF for the displacement vector from Eq.~\eqref{all
  terms}. The discrepancy between the immediate (un)folding limit and
dynamics in the target (un)folded structure (yellow and blue curves in
Fig.~\ref{fg4}b,e, respectively) is solely caused by the different
initial conditions whereas the propagator is in both cases
identical. Notably, this difference is small in the unfolding process
but substantial during folding. Several additional remarks are in
order. The ISF reflects dynamics
on a given spatial (i.e.\ $q$-value) scale and thus the projections
onto different relaxation eigenmodes play an important role (see also
e.g.\ \cite{faster}). The aforementioned discrepancies  are a result of initial
conditions (and starting conditions upon a jump) that put more weight
onto faster relaxation modes rendering the decay of the ISF faster
(this idea is corroborated in
Fig.~\ref{fg4}e). Note that relaxation in the folded state is faster than
in the unfolded and yet
the ISF in the folded state in Fig.~\ref{fg4}b decays the slowest.
This means that said differences are not only caused by a shorter
relaxation time (i.e.\ larger principal eigenvalue of the underlying
generator). 

For a comparison, in Fig.~\ref{fg4}c,f we also show the incoherent ISF
for the time-dependent mix, $F^{\rm mix}_{\rm u/f}(\vec
q,t)\equiv\mathrm{e}^{-k_{\rm u/f}t}\sum_{\alpha}F^{\alpha\alpha}_{\rm f/u}(\vec
q,t)+(1-\mathrm{e}^{-k_{\rm u/f}t})\sum_{\alpha}F^{\alpha\alpha}_{\rm u/f}(\vec
q,t)$, which represents the
trivial time-dependent approximation of the ISF for the unfolding and folding
processes, respectively.
In the case of unfolding the (trivial) mix
displays qualitative differences with respect to exact the two-state
(single jump) solution but the quantitative differences are rather small
(see Fig.~\ref{fg4}c). Conversely, in the case of folding the
mix-approximation $F^{\rm mix}_{\rm f}$  fails completely (see Fig.~\ref{fg4}f) due to the striking
dependence on the initial condition that is not included in the
(trivial) mix.  This clearly illustrates the importance of considering jump dynamics explicitly. 

Overall, our results suggest that scattering fingerprints may be a useful observable to probe
two-state dynamics even in the presence of non-trivial internal
structural relaxation, such as e.g.\ probed in
Ref.~\cite{Unfolding,Unfolding_2,Folding}.
A systematic study of the
relaxation at various values of the dimensionless quantities $k_{u,f}t_0$ and $t_0k/\xi$ may provide further details, which is,
however, beyond the scope of the present proof-of-principle
investigation. Moreover, one can easily generalize the model to
include internal friction \cite{Cheng2013} or hydrodynamic
interactions in the spirit of the Zimm model \cite{Zimm1956}. Finally,
one may also consider springs with a non-zero rest length in the context
of so-called Gaussian models (see e.g.~\cite{Lapolla_2021}).

\section{Conclusion}
Scattering experiments on polymerizing chains revealed pronounced
signatures of multi-state
dynamics \cite{Monkenbusch2016}. Moreover, observations of particle transport in complex environments such as the interior of living cells often reveal a non-Fickian (or non-Brownian) character
that may also display multi-state characteristics
\cite{Szabo_1982,Zwanzig_1992,Gated,Gated2,Emerging_JPA,Mirny,Koslover_2011,Sheinman_2012,Yamamoto,Kaerger,Grebenkov,Weiss2020}. 
Reliably
distinguishing between single- and multi-state dynamics 
remains challenging. To complement the 
existing approaches
we addressed scattering fingerprints of two-state dynamics -- the
intermediate scattering function and structure factor.
A combination of
theory, analysis of simple model systems and of 
experiments in living cells revealed the 
potential usefulness of these scattering fingerprints.

We addressed both, inert tracer-particle dynamics as well as dynamics
in systems with an internal structure and dynamics. In all examples
the ISF decays faster for increasing switching rates, which is
consistent with the idea of an additional relaxation channel enabling
a faster decay \cite{Granek1992,Stukalin2006}. However, this view contradicts
the
multi-state dynamics in
\cite{Monkenbusch2016}, where an empirical ansatz
assumed a mode-scission that leads to a slower decay of the ISF. 

The present results can be
extended to incorporate 
non-Markovian (i.e.\ non-exponential) waiting time statistics in the
respective states (see e.g.\ \cite{Networks,Krapf2019PT,Weron2017SR}), and may be relevant
and useful for digital Fourier microscopy (differential dynamic
microscopy and  Fourier imaging correlation spectroscopy) experiments
on complex particulate systems, as well as neutron (incl.\ spin-echo)
and X-ray scattering scattering probing structural and dynamical
properties of 
macromolecules, as soon as the dynamics displays two-state
transport.

\ack
We thank Maximilian Vossel for preparing Fig.~4a,d, and Adal Sabri for providing the data from Ref.~\cite{Weiss2020}. The financial support from the Studienstiftung des Deutschen Volkes (to
C.~D.), and the German Research Foundation (DFG) through the
\emph{Emmy Noether Program "GO 2762/1-1"} (to A.~G.) is gratefully acknowledged.\\

\appendix
\section{Derivation of equation (10)}\label{apSingwi}
Here we derive Eq.~\eqref{singwi series} for the Laplace-transformed
intermediate scattering function (ISF) of the two-state dynamics
following the Ref.~\cite{Singwi1960}. For this approach to hold,
memory in the process has to be erased at the instance of the jump and
the instantaneous position at the jump must not influence the
probability of subsequent displacements. These assumptions can arise
as a consequence of translation invariance of the dynamics in single
dynamical states, as in the examples of two-state diffusion and
FBM. However, translation invariance is not a necessary condition, as
e.g.\ in the case of the process studied in \cite{Singwi1960}. 

For a Markov jump process the probability to remain in a state for a
time $t$ is given by $\psi_j(t)=\rme^{-k_j t}$.  Recall that the van Hove
functions $G_j(\vec r,t)=\langle\delta(\vec r(t)-\vec r(0)-\vec
r)\rangle_j$ in the two individual states $j=1,2$ give the
probability of performing a displacement $\vec r$ in time $t$.
We now set $N=1$ and omit the indices $\alpha,\beta$ compared to
  Eq.~\eqref{vanHove-component}. For $N>1$ the assumption of memory erasure
  is rarely satisfied as a result of internal dynamics
  (see e.g.\ Sec.~\ref{unfolding}). If it is satisfied also for $N>1$, as e.g.\ in Sec.~\ref{dimerization} where $N=2$, then for
  $\alpha=\beta$ the approach does not change and Eq.~\eqref{singwi
    series} gives an equation for $\widehat{F}^{\alpha\alpha}(\vec
  q,s)$ that can subsequently be summed over $\alpha$. For
  $\alpha\neq\beta$ only the displacement probability prior to the
  first memory erasure is different, and the approach below may applied with
  slight modifications.

Following \cite{Singwi1960}, we write down the probability
$\mathcal{P}_n(\vec r,t)$ of a displacement $\vec r$ in time $t$
conditioned on a fixed number of $n$ jumps for $n=0,1,2$, given that
we start in state $1$ at $t=0$, and adopting the notation $G_j(\vec r,t)\psi_j(t)\equiv(G\psi)_j(\vec r,t)$,
\begin{align}
\mathcal{P}_0(\vec r,t)&=(G\psi)_1(\vec r,t)
\nonumber\\
\mathcal{P}_1(\vec r,t)&=\int\rmd^3 r_1\int_0^t\rmd t_1k_1(G\psi)_1(\vec r_1,t_1)(G\psi)_2(\vec r-\vec r_1,t-t_1)
\nonumber\\
\mathcal{P}_2(\vec r,t)&=\int\rmd^3 r_1\int\rmd^3 r_2\int_0^t\rmd t_2\int_0^{t_2}\rmd t_1k_1(G\psi)_1(\vec r_1,t_1)k_2(G\psi)_2(\vec r_2-\vec r_1,t_2-t_1)
\times\nonumber\\&\qquad\qquad\qquad\qquad\qquad\qquad\qquad
(G\psi)_1(\vec r-\vec r_2,t-t_2).
\end{align}
By the assumption of independence of the position at the time of the
jump, this has a convolution structure in space and we Fourier
transform $\vec r\to\vec q,\ \mathcal{P}\to\widetilde{\mathcal{P}}$ to obtain, e.g.\ for $n=2$,
\begin{align}
\widetilde{\mathcal{P}}_2(\vec q,t)&=\int_0^t\rmd t_2\int_0^{t_2}\rmd t_1k_1\widetilde{(G\psi)}_1(\vec q,t_1)k_2\widetilde{(G\psi)}_2(\vec q,t_2-t_1)\widetilde{(G\psi)}_1(\vec q,t-t_2).
\end{align}
Recalling that the intermediate scattering function (ISF) $F(\vec q,t)$
is the Fourier transform of the van Hove function $G(\vec r,t)$, and
noting the convolution structure (defined as
$[f*g](t)\equiv\int_0^t\rmd t'f(t')g(t-t')$) in time, we have 
\begin{align} 
\widetilde{\mathcal{P}}_2(\vec q,t)&=\int_0^t\rmd t_2\int_0^{t_2}\rmd t_1k_1(F\psi)_1(\vec q,t_1)k_2(F\psi)_2(\vec q,t_2-t_1)(F\psi)_1(\vec q,t-t_2)\nonumber\\
&=k_1k_2\int_0^t\rmd t_2(F\psi)_1(\vec q,t-t_2)\left[(F\psi)_1(\vec q,\cdot)*(F\psi)_2(\vec q,\cdot)\right](t_2)\nonumber\\
&=k_1k_2\left[(F\psi)_1*(F\psi)_1*(F\psi)_2\right](\vec q,t).
\end{align}
Taking the Laplace transform $t\to s$ gives
\begin{align}
\widehat{\mathcal{P}}_2(\vec q,s)=k_1k_2\left[\widehat{(F\psi)}_1(\vec q,s)\right]^2\widehat{(F\psi)}_2(\vec q,s).
\end{align}
This generalizes to all even $n=2m$ and odd $n=2m+1$ terms as
\begin{align}
\widehat{\mathcal{P}}_{2m+1}(\vec q,s)&=k_1^{m+1}k_2^m\left[\widehat{(F\psi)}_1(\vec q,s)\right]^{m+1}\left[\widehat{(F\psi)}_2(\vec q,s)\right]^{m+1},\nonumber\\
\widehat{\mathcal{P}}_{2m}(\vec q,s)&=k_1^mk_2^m\left[\widehat{(F\psi)}_1(\vec q,s)\right]^{m+1}\left[\widehat{(F\psi)}_2(\vec q,s)\right]^{m}.
\end{align} 
The Fourier-Laplace transform of the probability of a displacement in
the two-state dynamics that by the assumption of memory erasure and
independence of the jump position is the Laplace-transformed
intermediate scattering function $\widehat{F}(\vec q,s)$, is then
given by the geometric series 
\begin{align}
\widehat{F}(\vec q,s)&
=\sum_{n=0}^\infty\widehat{\mathcal{P}}_n(\vec q,s)\nonumber\\
&=\sum_{m=0}^\infty\left[\widehat{\mathcal{P}}_{2m}(\vec q,s)+\widehat{\mathcal{P}}_{2m+1}(\vec q,s)\right]\nonumber\\
&=\widehat{(F\psi)}_1(\vec q,s)\left[1+k_1\widehat{(F\psi)}_2(\vec q,s)\right]\sum_{m=0}^\infty\left[k_1k_2\widehat{(F\psi)}_1(\vec q,s)\widehat{(F\psi)}_2(\vec q,s)\right]^m\nonumber\\
&=\widehat{(F\psi)}_1(\vec q,s)\frac{1+k_1\widehat{(F\psi)}_2(\vec q,s)}{1-k_1k_2\widehat{(F\psi)}_1(\vec q,s)\widehat{(F\psi)}_2(\vec q,s)}.
\end{align}
Assuming initial occupations $p_j$ of the two states (not
necessarily $p_j=p_j^{\rm eq}$), and repeating the same treatment for
starting conditions in state 2 yields the result Eq.~\eqref{singwi series},
\begin{equation}
  \widehat{F}(\vec q,s)= \frac{p_1
    \widehat{(F\psi)}_1\left[1+k_1\widehat{(F\psi)}_2\right]+p_2
    \widehat{(F\psi)}_2\left[1+k_2\widehat{(F\psi)}_1\right]}{1-k_1k_2\widehat{(F\psi)}_1\widehat{(F\psi)}_2}. 
\end{equation}

\section{Two-state diffusion in the slow- and fast-switching limit}\label{apDiffusions} 
Here we consider the slow-switching limit of two-state diffusion in Sec.~\ref{twoch}, that is, the limit of small switching rates, $k_1+k_2\ll q^2D_{1,2}$. In this limit, Eq.~\eqref{mu} gives $\mu_+\approx q^2D_1$ and $\mu_-\approx q^2D_2$. Since $p_2=1-p_1$ we obtain $\phi(q)\approx p_2$ and thus Eq.~\eqref{diffusions-ISF} in the slow-switching limit results in the frozen mixture,
\begin{align}
F(q,t)\approx p_1\mathrm{e}^{-q^2D_1t}+p_2\mathrm{e}^{-q^2D_2t}=F_{\rm mix}(q,t).
\end{align} 

Now consider the fast-switching limit, $k_1+k_2\gg q^2D_{1,2}$ with a
finite ratio $k_1/k_2=p_2^{\rm eq}/p_1^{\rm eq}$ (when this ratio is
zero or infinite we recover single-state dynamics).  We introduce the notation $k\equiv (k_1+k_2)/2$, $\kappa\equiv (k_1-k_2)/2$, $\bar{d}\equiv q^2(D_1+D_2)/2$ and $\Delta\equiv q^2(D_2-D_1)/2$. Then Eq.~\eqref{mu} reads
\begin{align}
        \mu_{1,2}&
        =\bar{d}+k\pm\sqrt{(\Delta+\kappa)^2+k^{2}-\kappa^{2}}\nonumber\\
        &=\bar{d}+k\pm\sqrt{\Delta^{2}+2\Delta\kappa+k^{2}}.
\end{align}
In the limit of fast jumps $k\gg\bar{d}>\abs{\Delta}$ we have
\begin{align}
        \mu_{1,2}&=\bar{d}+k\pm k\sqrt{1+\frac{2\Delta\kappa+\Delta^{2}}{k^{2}}}
        \approx\bar{d}+k\left[1\pm\left(1+\frac{2\Delta\kappa+\Delta^{2}}{2k^{2}}\right )\right ],\nonumber\\
		\mu_1&\approx\bar{d}-\frac{2\Delta\kappa+\Delta^{2}}{2k}\approx\bar{d}-\frac{\kappa}{k}\Delta,\quad \mu_2\approx\bar{d}+2k+\frac{2\Delta\kappa+\Delta^{2}}{2k}\approx 2k\overset{k\to\infty}{\longrightarrow}\infty.
\end{align}
Using Eq.~\eqref{p_i}, we have $k_1=2p^{\rm eq}_2k$ and $k_2=2p^{\rm
  eq}_1k$ and arrive at
\begin{align}
\mu_1&\approx\bar{d}-\frac{\kappa}{k}\Delta=\bar{d}+\Delta\frac{p^{\rm eq}_2-p^{\rm eq}_1}{p^{\rm eq}_2+p^{\rm eq}_1}
=q^2(p^{\rm eq}_1D_1+p^{\rm eq}_2D_2).
\end{align}
Thus, the exponential with rate $\mu_1$ in Eq.~\eqref{diffusions-ISF} gives an exponential with effective diffusion constant $D_{\rm eff}\equiv p^{\rm eq}_1D_1+p^{\rm eq}_2D_2$, while the second exponential with rate $\mu_2\approx 2k$ immediately decays in the limit of large $k$. This proves that for large $k$ the ISF of two-state diffusion approaches effective diffusion, i.e.\ in Fig.~\ref{fg1}a the dashed lines approach the yellow line.

\section{Laplace transforms for fractional Brownian motion}\label{apFBM}
We consider fractional Brownian motion (FBM) in $d$ dimensions with mean squared displacement
\begin{align}
    \left\langle [\vec r(t)-\vec r(0)]^2\right\rangle=2d C_j t^{\alpha_j}.
\end{align}
Here, the two different states are characterized by different generalized diffusion constants $C_1$ and $C_2$ and/or different ``anomalous'' exponents $\alpha_1$ and $\alpha_2$. We still consider Markov jumps and by Gaussianity we have the Fourier-transformed displacement probabilities in a state
\begin{align}
    F_j(\vec q,t)\psi_j(t)&=\exp(-[q^2 C_j t^{\alpha_j}+k_j]\,t).
    \label{FBM displacement probability}
\end{align}
Assuming that each jump (i.e.\ change of state) erases memory the structure factor for the two-state dynamics follows from Eq.~\eqref{singwi series} which requires taking the Laplace transform of Eq.~\eqref{FBM displacement probability}. For our purposes we consider subdiffusive FBM with $\alpha\in(0,1)$ (for $\alpha=1$ see Eq.~\eqref{two}). The Laplace transform of Eq.~\eqref{FBM displacement probability} can of course be performed numerically\cite{Wuttke2012A}, and for any rational $\alpha\in(0,1)$ it can be expressed in terms of Meijer G-functions, using that for 
two positive integers $l<m$ \cite{Prudnikov1992}
\begin{align}
\widehat{[\rme^{-at^{l/m}}]}(s)=\frac{\sqrt{ml}}{s(2\pi)^{(m+l)/2-1}}G^{m,l}_{l,m}\left(\frac{a^ml^l}{m^mp^l}\bigg|\genfrac{}{}{0pt}{}{\Delta(l,0)}{\Delta(m,0)}\right).\label{Meijer} 
\end{align}
For the special case $\alpha=0.5$ the Laplace transform of Eq.~\eqref{FBM displacement probability} follows from \cite{Prudnikov1992}
\begin{align}
\widehat{[\rme^{-a\sqrt{t}}]}(s)=\frac{1}{s}-\frac{a}{2}\sqrt{\frac{\pi}{s^3}}{\rm erfcx}\left(\frac{a}{2\sqrt{s}}\right),
\end{align}
where $\text{erfcx}(x)=\exp(x^2)(1-\text{erf}(x))$ and erf denotes the error function, and therefore
\begin{align}
    \widehat{(F\psi)}_j(\vec q,s)=\frac{1}{s+k_j}-\frac{\sqrt{\pi}q^2 C_j}{2(s+k_j)^{3/2}}\text{erfcx}\left(\frac{q^2 C_j}{2\sqrt{s+k_j}}\right).
\end{align}
Inserting $s=-i\om$ and taking the real part as in
Eq.~\eqref{diffusions-DSF}, we obtain the DSF. Numerically
transforming back to the time-domain in turn yields the ISF.

For $\alpha\ne 0.5$, instead of using Meijer-G-functions or numerical solutions we can perform a series expansion \cite{Wuttke2012A}. We expand $\rme^{-ct^\alpha}$ around $t=0$
\begin{align}
    \int_0^\infty \rme^{-st}\rme^{-ct^\alpha}\rmd t&=\sum_{n=0}^\infty \frac{(-1)^n c^{n}}{n!}\int_0^\infty \rme^{-st} t^{n\alpha}\rmd t\nonumber\\
    &=\sum_{n=0}^\infty \frac{(-1)^n \Gamma(\alpha n+1)}{n!}c^{n}s^{-(\alpha n+1)},
    \label{FBM laplace series}
\end{align}
where the sum and integral commute since the sum is bounded by $\rme^{-c t^\alpha},\ c>0$ which is integrable for $\alpha<1$.
A series for the Laplace transform of Eq.~\eqref{FBM displacement probability} is obtained by shifting $s\mapsto s+k_j$ and using $c=q^2C_j$. This approximation works particularly well for or small $q$, not too small $s,k$ and $\alpha<1$ not too close to $1$.

The complementary expansion for small $s$ reads
\begin{align}
    \int_0^\infty \rme^{-st}\rme^{-ct^\alpha}\rmd t&
    \approx \sum_{n=0}^N \frac{(-1)^n s^{n}}{n!}\int_0^\infty t^n \rme^{-ct^\alpha}\rmd t\nonumber\\
    &=\frac{1}{\alpha}\sum_{n=0}^N \frac{(-1)^n}{n!}\Gamma\left(\frac{n+1}{\alpha}\right)c^{-\frac{n+1}{\alpha}}s^n.
    \label{FBM laplace series - small s}
\end{align}
However, note that for $\alpha<1$ this series is only asymptotically convergent \cite{Wuttke2012A}, and therefore has to be truncated at some finite $N$ to be able to swap the integration and summation. We use this expansion in the very small $s$ regime in cases where the expansion \eqref{FBM laplace series} does not suffice. 

\begin{figure} \centering
\includegraphics[scale=1]{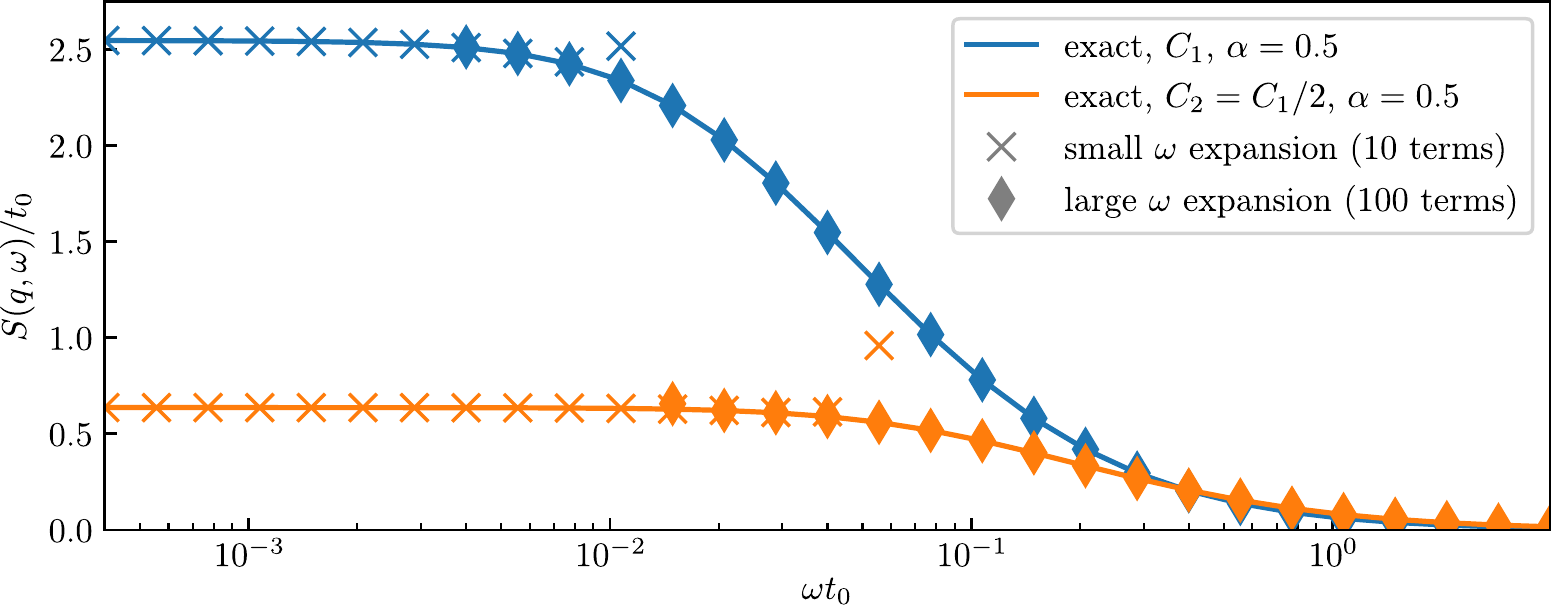}
\caption{\label{fgAp1} Illustration of the asymptotics for the DSF as introduced in Eq.~\eqref{DSF}. Time is measured in units of $t_0=1/(C_1^2q^4)$. The large-$\omega$ series \eqref{FBM laplace series} and the small-$\om$ expansion \eqref{FBM laplace series - small s} together give a very good approximation to the exact DSF, i.e.\ the Fourier transform of Eq.~\eqref{FBM displacement probability} with $k=0$.}
\end{figure}

The two asymptotic results combined yield a very good approximation as
illustrated for the DSF of a FBM in Fig.~\ref{fgAp1} (recall that the
DSF follows from the Laplace transform of Eq.~\eqref{FBM displacement
  probability} for $k_j=0$ by setting $s=-\rmi\om$ and taking the real
part as in Eq.~\eqref{DSF}). Combining the two expansions allows for
efficient computation of $(\widehat{F\psi})(q,s)$ and $S(q,\omega)$
over the whole $\omega$ regime for general $\alpha\in(0,1)$, which in turn allows via Eq.~\eqref{singwi series} to deduce the two-state ISF.

\section{Simulation details}\label{apSim}
The experimental data shown in Fig.~\ref{fg2} consists of 200
two-dimensional trajectories which by assuming isotropy are treated as
400 one-dimensional realizations of the two-state FBM. The
experimental ISF is obtained by averaging $e^{-iq [r(t)-r(0)]}$ over
the 400 realizations. Although the ISF of the model is deterministic,
the experimental ISF obtained from a finite number of realizations
fluctuates around the deterministic ISF. To estimate these
fluctuations we performed simulations of $20,000=500\times 400$
two-state FBM trajectories with the parameters obtained by the fit.
The standard deviation in each point of the ISF averaged over sets of 400 trajectories is shown in Fig.~\ref{fg2}. Moreover, the MSD and its fluctuations upon averaging over 400 realizations are obtained from the simulations and shown in Fig.~\ref{fg2}.

Simulations were performed by drawing exponentially distributed
waiting times for the state switching and simulating FBM
trajectories between the switching times. The FBM simulations were
performed using the Davies-Harte-algorithm \cite{Davies1987B}
using the "fbm" python package available on PyPI. 
  
\section{Details on dimerization with an internal mode}\label{apDimer}
The ISF of the dimer is the characteristic function of a Gaussian with displacement given by Eq.~\eqref{displacement single process} for $N=2$ for which \begin{align}A=\begin{bmatrix}1&-1\\ -1&1\end{bmatrix},\qquad Q=\frac{1}{\sqrt{2}}\begin{bmatrix}1&-1\\1&1\end{bmatrix}.\end{align} With $D_{\rm COM}=D_2=D/2$, $V(\infty)=3D/a$ and $\psi_2(t)=\rme^{-k_2t}$ we obtain
\begin{align}
(F\psi)_2(q,t)&=\exp\left[-q^2D_2t-k_2 t-\frac{q^2D}{2a}\left(1-\rme^{-at}\right)\right]\nonumber\\
    &=\rme^{-\frac{q^2 D}{2a}}\rme^{-(q^2D_2+k_2)t}\exp\left(\frac{q^2 D}{2a}\rme^{-at}\right),
\end{align}
where for convenience we have assumed equilibrium initial conditions of the dimer.

The Laplace transform can be conveniently carried out via a series
expansion (the order of summation and integration can be interchanged due to dominated convergence, i.e.\ since the series is bounded from above by the full expression),
\begin{align}
\widehat{(F\psi)}_2(q,s)&=\rme^{-\frac{q^2D}{2a}}\int_0^\infty\rmd t\ \rme^{-(s+q^2D_2+k_2)t}\exp\left(\frac{q^2 D}{2a}\rme^{-at}\right)\nonumber\\
&=\rme^{-\frac{q^2 D}{2a}}\sum_{n=0}^\infty\frac{(\frac{q^2D}{2a})^{n}}{n!}\int_0^\infty \rmd t\ \rme^{-(s+q^2D_2+k_2+na)t}\nonumber\\
&=\rme^{-\frac{q^2D}{2a}}\sum_{n=0}^\infty\,\frac{(\frac{q^2D}{2a})^{n}}{n!}\frac{1}{s+q^2D_2+k_2+na}.
\end{align}
The series expansion converges very well, especially for small $q$ and
large $t$ (or small $\om$). Note that there exists a closed form
expression of $\widehat{(F\psi)}_2(q,s)$ in terms
Kummer's hypergeometric function \cite{Prudnikov1992} (or in terms of the incomplete Gamma
function  \cite{Kneller2005}) as
in Eq.~\eqref{dimer}.

In the limit of fast jumps $k_1,k_2\gg q^2 D$ the time
spent in a single state prior to a jump becomes very small, $at\ll 1$, for which we expand
\begin{align}
    (F\psi)_2(\vec q,t)&=\rme^{-\frac{q^2 D}{2a}}\rme^{-(q^2D_2+k_2)t}\exp\left(\frac{q^2 D}{2a}\rme^{-at}\right)\nonumber\\
    &\overset{t\to 0}{\approx}\rme^{-\frac{q^2D}{2a}}\rme^{-(q^2D_2+k_2)t}\rme^{\frac{q^2D}{2a}(1-at)}\nonumber\\
    &=\rme^{-(q^2D+k_2)t}\nonumber\\
    &=F_1(\vec q,t)\psi_2(t).
\end{align}
Therefore, for $k_1,k_2\gg q^2 D$ the two-state ISF converges to the
simple monomer diffusion, 
i.e.\ the dashed lines approach the yellow line in Fig.~\ref{fg3}b.

\section{Two-state dynamics with internal structure/dynamics -- irreversible protein (un)folding}\label{apENM}
Here we derive Eqs.~\eqref{displacement single process}-\eqref{all terms} using normal mode analysis as introduced in
Sec.~\ref{unfolding}, i.e.\ in the normal coordinates $\mathbf{X}^{\rm f/u}=(\Q^{\rm f/u})^T\mathbf{R}$ with
$(\Q^{\rm f/u})^T\A^{\rm f/u}\Q^{\rm f/u}={\rm diag}(a_\alpha^{\rm
  f/u})_{\alpha=1,\ldots,N}$. The drift matrix $\A^{\rm f/u}$ can be considered as $N\times N$-dimensional since the model assumes isotropy in all three dimensions and thus each eigenspace of the $3N\times 3N$ drift matrix is threefold degenerate. Note that $(Q^{\rm f/u})^{-1}=(Q^{\rm
  f/u})^{T}$, 
  and in normal coordinates (omitting
index $\rm f/u$) we have
$\mathrm{d}\vec{X}^{\alpha}_t=-a_\alpha\vec{X}^{\alpha}_t{\rm
  d}t+\sqrt{2D}{\rm d}\vec{W}^{\alpha}_t$.

We choose the first mode $\alpha=1$ to correspond to the zero eigenvalue, i.e.\ $a_1=0$ and thus $\vec X^1_t$ evolves as a free diffusion, capturing to the
center-of-mass motion with $Q^{\rm f/u}_{\alpha1}=N^{-1/2}$ and $D_{\rm COM}=D/N$. The position at time $t=0$ of the center
of mass cancels out in all displacements and only differences between
time $0$ and $t$ enter. Thus, whenever we speak of the equilibrium initial conditions we refer to the equilibrium of modes $\vec X^\alpha,\ \alpha>1$, and the initial condition of the $\vec X^1$-mode is irrelevant.

The $\vec X^\alpha$ modes with $\alpha>1$ have $a_\alpha>0$ and are thus described by an Ornstein-Uhlenbeck process \cite{Ornstein1930}, for which we obtain from the stochastic differential equation that for any $\tau\in[0,t]$
\begin{align}
\rmd\left\langle\vec X^\alpha_{t'}\cdot\vec  X^\beta_{0}\right\rangle&=-a_\alpha\left\langle\vec X^\alpha_{t'}\cdot\vec  X^\beta_{0}\right\rangle\rmd t'\quad
\Rightarrow\quad \left\langle\vec X^\alpha_t\cdot\vec  X^\beta_0\right\rangle&=\rme^{-a_\alpha(t-\tau)}\left\langle\vec X^\alpha_\tau\cdot\vec  X^\beta_0\right\rangle,
\label{OUP correlation}
\end{align}
and using It\^o's lemma as well as the shorthand notation $V^\alpha(t)=\frac{3D}{a_\alpha}(1-\rme^{-2a_\alpha t})$,
\begin{align}
\rmd\left\langle\vec X^\alpha_{t'}\cdot\vec X^\beta_{t'}\right\rangle&=\left\langle\rmd \vec X^\alpha_{t'}\cdot\vec X^\beta_{t'}\right\rangle+\left\langle\vec X^\alpha_{t'}\cdot\rmd\vec X^\beta_{t'}\right\rangle+\frac{1}{2}\left\langle\rmd \vec X^\alpha_{t'}\cdot\rmd\vec X^\beta_{t'}\right\rangle
\nonumber\\
&=-(a_\alpha+a_\beta)\left\langle\vec X^\alpha_{t'}\cdot\vec X^\beta_{t'}\right\rangle\rmd t'+3D\delta_{\alpha\beta}\rmd t'
\nonumber\\
\Rightarrow\quad\left\langle\vec X^\alpha_t\cdot\vec X^\beta_t\right\rangle&=\left\langle\vec X^\alpha_\tau\cdot\vec X^\beta_\tau\right\rangle\rme^{-(a_\alpha+a_\beta)(t-\tau)}+\delta_{\alpha\beta}V^\alpha(t-\tau)
,
\label{OUP variance}
\end{align}
and thus for equilibrium initial conditions $\left\langle\vec X^\alpha_t\cdot\vec
X^\beta_t\right\rangle_{\rm
  eq}=\delta_{\alpha\beta}V^\alpha(\infty)\equiv \delta_{\alpha\beta}V^\alpha$.

Since the Ornstein-Uhlenbeck process with a Gaussian initial condition is a
Gaussian process,  $F^{\alpha\beta}_{\rm u}(\vec q,t)$ and $F^{\alpha\beta}_\utof(\vec q,t;\tau)$ are characteristic functions, of zero mean Gaussian displacement vectors. For any real Gaussian
stochastic process $Y_{t},\,0\le t\le T$ the characteristic function
of the joint probability distribution of the random variables
$Y_{t_1}\ldots Y_{t_n}$ with $0\le t_1\ldots t_n\le T$ ($n<\infty$)
is given by $\phi(q_1\ldots
q_n)=\exp(i\sum_{k=1}^n\langle Y_{t_k}\rangle
q_k-\sum_{k,l=1}^n\langle[Y_{t_k}-\langle
  Y_{t_k}\rangle][Y_{t_l}-\langle Y_{t_l}\rangle]\rangle q_kq_l/2)$,
which naturally generalizes to $d$-dimensional Gaussian stochastic
process with vector values $\vec{Y}_t=(Y^1_{t},\ldots,Y^d_t)$
($d<\infty$) \cite{Doob}. In our case we have $\vec{Y}_{t_k}=
\vec{R}^{\alpha}_{t_k}-\vec{R}^{\beta}_{t_{k-1}}$, $n=1$, and $\langle
\vec{Y}_{t_k}\rangle=0,\,\forall t_k$ because all involved Gaussian
distributions are centered at zero, and the additional factor of $1/3$
in $q^2/6$ is due to isotropy. Therefore, $F^{\alpha\beta}_{{\rm u}/\utauf}(\vec q,t)={\rm exp}{[-q^2\langle(\vec R^\alpha_t-\vec R^\beta_0)^2\rangle_{{\rm u}/\utauf}/6]}$ and we only need to compute the squared displacements $\langle(\vec R^\alpha_t-\vec R^\beta_0)^2\rangle_{{\rm u}/\utauf}$ to obtain the ISFs.

We first consider the single-state process, say in the unfolded
state $\rm u$, with equilibrium initial conditions and now derive the
result Eq.~\eqref{displacement single process}. Using normal
coordinates $\f R=\Q^{\rm u}\f X^{\rm u}$ (from now on omit index $\rm
u$ in $\f X^{\rm u}$), the independence of modes, and that the mean
value vanishes, we obtain
\begin{align}
\left\langle\left[\vec R^\alpha_t-\vec R^\beta_0\right]^2\right\rangle_{\rm u}
&=\left\langle\left[\sum_{\gamma=1}^N\left (Q^{\rm u}_{\alpha\gamma}\vec X^\gamma_t-Q^{\rm u}_{\beta\gamma}\vec X^\gamma_0\right )\right]^2\right\rangle_{\rm u}\nonumber\\
&=\sum_{\gamma=1}^N\left\langle\left (Q^{\rm u}_{\alpha\gamma}\vec X^\gamma_t-Q^{\rm u}_{\beta\gamma}\vec X^\gamma_0\right )^2\right\rangle_{\rm u}.
\end{align}
The center-of-mass motion with $D_{\rm COM}=D/N$ is obtained from $Q_{\alpha1}^{\rm u}=Q_{\beta1}^{\rm u}=N^{-1/2}$. Using Eqs.~\eqref{OUP correlation}-\eqref{OUP variance} we directly obtain the result Eq.~\eqref{displacement single process},
\begin{align}
\left\langle\left(\vec R^\alpha_t-\vec R^\beta_0\right)^2\right\rangle_{\rm u}
&=6D_{\rm COM}t+\sum_{\gamma=2}^NV_{\rm u}^\gamma\left[(Q_{\alpha\gamma}^{\rm u})^2+(Q_{\beta\gamma}^{\rm u})^2-2Q^{\rm u}_{\alpha\gamma}Q^{\rm u}_{\beta\gamma}\rme^{-a_\gamma^{\rm u}t}\right].
\end{align}

We now consider the process with a single jump at a fixed time
$\tau<t$, i.e.\ we start in the equilibrium of the unfolded $\rm
u$-state and propagate in the $\rm u$-state from time $0$ to $\tau$
and in the folded $\rm f$-state from time $\tau$ to $t$. Averages with
respect to this process will be denoted by
$\langle\cdot\rangle_\utauf$\,. Within the Gaussian network model the
center-of-mass motion is independent of the network structure and we
obtain similar to above (but now the modes are mixed due to $\Q^{\rm u}\ne\Q^{\rm f}$ and thus no longer decouple),
\begin{align}
&\left\langle\left(\vec R^\alpha_t-\vec R^\beta_0\right)^2\right\rangle_{\utauf}
=6D_{\rm COM}t+\sum_{\gamma,\nu=2}^N\Big[Q_{\alpha\gamma}^{\rm u}Q_{\alpha\nu}^{\rm u}\left\langle(\f X_t^{\rm f})_\gamma\cdot(\f X_t^{\rm f})_\nu\right\rangle_\utauf
\nonumber\\&+Q_{\beta\gamma}^{\rm u}Q_{\beta\nu}^{\rm u}\left\langle(\f X_0^{\rm f})_\gamma\cdot(\f X_0^{\rm f})_\nu\right\rangle_\utauf-2Q^{\rm u}_{\alpha\gamma}Q^{\rm u}_{\beta\nu}\left\langle(\f X_t^{\rm f})_\gamma\cdot(\f X_0^{\rm f})_\nu\right\rangle_\utauf\Big].
\label{intermediate1}
\end{align}
Now we determine the three terms $\left\langle(\f X_t^{\rm f})_\gamma\cdot(\f X_t^{\rm f})_\nu\right\rangle_\utauf$, $\left\langle(\f X_0^{\rm f})_\gamma\cdot(\f X_0^{\rm f})_\nu\right\rangle_\utauf$ and $\left\langle(\f X_t^{\rm f})_\gamma\cdot(\f X_0^{\rm f})_\nu\right\rangle_\utauf$.
The first term resembles $\left\langle(\f X_t^{\rm f})_\gamma(\f
X_t^{\rm f})_\nu\right\rangle_{\rm f}=\delta_	{\gamma\nu}V_{\rm
  f}^\gamma$ with the difference that we do not start in the $\rm f$-equilibrium but propagate the $\rm f$-process from the $\rm u$-equilibrium from time $\tau$ to $t$. In particular, modes are mixed since correlations at time $\tau$ are only diagonal in $\rm u$-normal modes $\f X^{\rm u}$ but not in $\f X^{\rm f}$. Using Eq.~\eqref{OUP variance} we have
\begin{align}
\left\langle(\f X_t^{\rm f})_\gamma\cdot(\f X_t^{\rm f})_\nu\right\rangle_\utauf&=
\left\langle(\f X_\tau^{\rm f})_\gamma\cdot(\f X_\tau^{\rm f})_\nu\right\rangle_\utauf\rme^{-(a_\gamma+a_\nu)(t-\tau)}+\delta_{\gamma\nu}V^\gamma(t-\tau).
\label{intermediate2}
\end{align}
Since we start in $\rm u$-equilibrium and propagate in the $\rm u$-process until time $\tau$, we complete the calculation of the first term by calculating the second term, by expressing $\rm f$-modes in terms of $\rm u$-modes. Using the notation $\Q^{{\rm u}\to\rm f}\equiv(\Q^{\rm f})^T\Q^{\rm u}$ (s.t.\ as in $\f R=\Q^{\rm u}\f X^{\rm u}=\Q^{\rm f}\f X^{\rm f}$ we have $\f X^{\rm f}=\Q^{{\rm u}\to\rm f}\f X^{\rm u}$),
\begin{align}
\left\langle(\f X_\tau^{\rm f})_\gamma\cdot(\f X_\tau^{\rm f})_\nu\right\rangle_\utauf&=\left\langle(\f X_0^{\rm f})_\gamma\cdot(\f X_0^{\rm f})_\nu\right\rangle_\utauf
=\left\langle(\f X_0^{\rm f})_\gamma\cdot(\f X_0^{\rm f})_\nu\right\rangle_{\rm u}\nonumber\\
&=\sum_{\omega=2}^NQ^{{\rm u}\to\rm f}_{\gamma\omega}Q^{{\rm u}\to\rm f}_{\nu\omega}V_{\rm u}^\omega.
\label{intermediate3}
\end{align}
For the third term we employ Eq.~\eqref{OUP correlation} for the $\rm f$-process from time $\tau$ to $t$ and for the $\rm u$-process from $0$ to $\tau$,
\begin{align}
\left\langle(\f X_t^{\rm f})_\gamma\cdot(\f X_0^{\rm f})_\nu\right\rangle_\utauf&=\rme^{-a_\gamma^{\rm f}(t-\tau)}\left\langle(\f X_\tau^{\rm f})_\gamma\cdot(\f X_0^{\rm f})_\nu\right\rangle_\utauf\nonumber\\
&=\rme^{-a_\gamma^{\rm f}(t-\tau)}\sum_{\omega,\omega'=2}^NQ_{\gamma\omega}^{{\rm u}\to\rm f}Q_{\nu\omega'}^{{\rm u}\to\rm f}\left\langle(\f X_\tau^{\rm u})_\omega\cdot(\f X_0^{\rm u})_{\omega'}\right\rangle_{\rm u}\nonumber\\
&=\rme^{-a_\gamma^{\rm f}(t-\tau)}\sum_{\omega=2}^NQ_{\gamma\omega}^{{\rm u}\to\rm f}Q_{\nu\omega}^{{\rm u}\to\rm f}V^\omega_{\rm u}\rme^{-a^{\rm u}_\omega\tau}.
\label{intermediate4}
\end{align}
Combining Eqs.~\eqref{intermediate1}-\eqref{intermediate4} and using $\sum_{\gamma\nu}C_\gamma C_\nu=(\sum_\gamma C_\gamma)^2$ and for the last term $\Q^{\rm f}\Q^{{\rm u}\to\rm f}=\Q^{\rm u}$ we arrive at the result Eq.~\eqref{all terms},
\begin{align}
&\left\langle\left(\vec R^\alpha_t -\vec R^\beta_0\right)^2\right\rangle_\utauf
=6D_{\rm COM}t+\sum_{\gamma=2}^N(Q^{\rm u}_{\beta\gamma})^2V_{\rm u}^{\gamma}+\sum_{\gamma=2}^N(Q^{\rm f}_{\alpha\gamma})^2V_{\rm f}^{\gamma}(t-\tau)\nonumber\\
&+\sum_{\omega=2}^NV_{\rm u}^{\omega}\left[\sum_{\gamma=2}^N Q^{\rm f}_{\alpha\gamma}Q^{{\rm u}\to\rm f}_{\gamma\omega}\rme^{-a^{\rm f}_\gamma(t-\tau)}\right]^2-2\sum_{\omega,\gamma=2}^NQ^{\rm f}_{\alpha\omega}Q^{{\rm u}\to\rm f}_{\omega\gamma}Q^{\rm u}_{\beta\gamma}V_{\rm u}^{\gamma}\rme^{-a^{\rm f}_\omega(t-\tau)-a^{\rm u}_\gamma\tau}.
\end{align}

Finally, for completeness we here show the
Kirchoff
  matrix $\A^{\rm f}$ corresponding to the Trp-Cage Miniprotein Construct TC5b (PDB: 1L2Y) 
\setcounter{MaxMatrixCols}{20}
\begin{align}
\A^{\rm f}=\frac{k}{\xi}
\tiny{
\begin{bmatrix} 
6& -1& -1& -1& -1& -1& -1&  0&  0&  0&  0&  0&  0&  0&  0&  0& 0&  0&  0&  0\\
       -1&  9& -1& -1& -1& -1& -1&  0&  0&  0&  0&  0&  0&  0&  0&  0&         0& -1& -1& -1\\
       -1& -1& 11& -1& -1& -1& -1& -1&  0&  0& -1&  0&  0&  0&  0&  0&         0& -1& -1& -1\\
       -1& -1& -1& 10& -1& -1& -1& -1& -1&  0& -1&  0&  0&  0&  0&  0&         0&  0& -1&  0\\
       -1& -1& -1& -1& 11& -1& -1& -1& -1& -1& -1&  0&  0&  0&  0&  0&         0&  0& -1&  0\\
       -1& -1& -1& -1& -1& 16& -1& -1& -1& -1& -1& -1&  0& -1&  0& -1&        -1& -1& -1&  0\\
       -1& -1& -1& -1& -1& -1& 14& -1& -1& -1& -1& -1& -1& -1&  0&  0&         0& -1&  0&  0\\
        0&  0& -1& -1& -1& -1& -1&  9& -1& -1& -1&  0&  0& -1&  0&  0&         0&  0&  0&  0\\
        0&  0&  0& -1& -1& -1& -1& -1& 12& -1& -1& -1& -1& -1& -1& -1&         0&  0&  0&  0\\
        0&  0&  0&  0& -1& -1& -1& -1& -1& 11& -1& -1& -1& -1& -1& -1&         0&  0&  0&  0\\
        0&  0& -1& -1& -1& -1& -1& -1& -1& -1& 15& -1& -1& -1& -1& -1&        -1& -1&  0&  0\\
        0&  0&  0&  0&  0& -1& -1&  0& -1& -1& -1& 11& -1& -1& -1& -1&        -1& -1&  0&  0\\
        0&  0&  0&  0&  0&  0& -1&  0& -1& -1& -1& -1&  9& -1& -1& -1&        -1&  0&  0&  0\\
        0&  0&  0&  0&  0& -1& -1& -1& -1& -1& -1& -1& -1& 11& -1& -1&        -1&  0&  0&  0\\
        0&  0&  0&  0&  0&  0&  0&  0& -1& -1& -1& -1& -1& -1&  8& -1&        -1&  0&  0&  0\\
        0&  0&  0&  0&  0& -1&  0&  0& -1& -1& -1& -1& -1& -1& -1& 11&        -1& -1& -1&  0\\
        0&  0&  0&  0&  0& -1&  0&  0&  0&  0& -1& -1& -1& -1& -1& -1&        10& -1& -1& -1\\
        0& -1& -1&  0&  0& -1& -1&  0&  0&  0& -1& -1&  0&  0&  0& -1&        -1& 10& -1& -1\\
        0& -1& -1& -1& -1& -1&  0&  0&  0&  0&  0&  0&  0&  0&  0& -1&        -1& -1&  9& -1\\
        0& -1& -1&  0&  0&  0&  0&  0&  0&  0&  0&  0&  0&  0&  0&  0&
        -1& -1& -1&  5
\end{bmatrix}},
\end{align}
and  the $20\times 20$ tridiagonal Kirchoff matrix $A^{\rm u}$ for the unfolded Rouse chain
\begin{align}
\A^{\rm u}=\frac{k}{\xi}\begin{bmatrix}1&-1&&&&&\\-1&2&-1&&&0&\\&-1&2&-1&&&\\&&&\dots&\dots&&\\&&&&\dots&\dots&\\&0&&&-1&2&-1\\&&&&&-1&1\end{bmatrix}.
\end{align}

\section*{References}

\bibliography{bib_rv.bib}

\end{document}